\newcommand{\be}{\begin{equation}}
\newcommand{\ee}{\end{equation}}
\newcommand{\bea}{\begin{eqnarray}}
\newcommand{\eea}{\end{eqnarray}}
\newcommand{\ba}[1]{\begin{array}{#1}}
\newcommand{\ea}{\end{array}}
\newcommand{\ket}[1]{|#1\rangle}
\newcommand{\bra}[1]{\langle#1|}

\documentclass[pra,aps,showpacs,twocolumn]{revtex4-1}
\usepackage{times}
\usepackage{amssymb}
\usepackage{amsmath}
\usepackage{mathrsfs}
\usepackage{graphicx}
\usepackage{epsfig}
\usepackage{dcolumn}
\usepackage{color}
\usepackage{bm}

\DeclareMathOperator{\T}{\mathcal{T}}

\begin{document}

\title{Photon Scattering from a System of Multi-Level Quantum Emitters. II. Application to Emitters Coupled to a 1D Waveguide}
\author{Sumanta Das$^{1}$, Vincent E. Elfving$^{1}$, Florentin Reiter$^{2}$, and Anders S. S\o rensen$^{1}$}
\affiliation{$^{1}$Niels Bohr Institute, University of Copenhagen, Blegdamsvej 17, 2100 Copenhagen \O, Denmark\\
$^{2}$ Department of Physics, Harvard University, Cambridge, MA 02138, USA}
\date{\today}
\begin{abstract}
In a preceding paper we introduced a formalism to study the scattering of low intensity fields from a system of multi-level emitters embedded in a $3$D dielectric medium. Here we show how this photon-scattering relation can be used to analyze the scattering of single photons and weak coherent states from any generic multi-level quantum emitter coupled to a $1$D  waveguide. The reduction of the photon-scattering relation to $1$D waveguides provides for the first time a direct solution of the scattering problem involving low intensity fields in the waveguide QED regime. To show how our formalism works, we consider examples of multi-level emitters and evaluate the transmitted and reflected field amplitude. Furthermore, we extend our study to include the dynamical response of the emitters for scattering of a weak coherent photon pulse. As our photon-scattering relation is based on the Heisenberg picture, it is quite useful for problems involving photo-detection in the waveguide architecture. We show this by considering a specific problem of state generation by photo-detection in a multi-level emitter, where our formalism exhibits its full potential. Since the considered emitters are generic, the $1$D results apply to a plethora of physical systems like atoms, ions, quantum dots, superconducting qubits, and nitrogen-vacancy centers coupled to a $1$D waveguide or transmission line.
\end{abstract}

\maketitle

\section{Introduction}
Efficient light-matter interfaces at the few to single-photon level are crucial for quantum information processing and future quantum technologies \cite{Kim08,Tey08, Brien09, Chang07, Hwang09}. Traditionally, such interfaces have been pursued with atoms coupled to a single mode of an optical cavity with a high Q factors, in the regime of cavity quantum electrodynamics (QEDs) \cite{Haroche_rmp}. The strong confinement of light in optical cavities, however, also poses a limitation to their integration into quantum networks, which relies on the efficient out-coupling of light \cite{Rempe_rmp}. As such, currently a wide variety of physical systems are being studied where one achieves good light-matter interface, which can be integrated in future with opto-electronics \cite{Kurt20, Brouri20, Yuan05, Shields07, Shen_prl07, Houck07, Fu08, Rebic09, Shi09, Babi10, Liew10, Shi11, Bamba11, Maju12, Pey12, Loo13, Shi13, Baur14, Tiecke14, Giesz16}. Among these, waveguides coupled to quantum emitters have turned out to be a viable alternative \cite{Peter_rmp}. 

The study of photon scattering in waveguides traditionally considers an emitter either coupled to a continuous set of freely propagating waveguide modes or coupled to a discrete set of modes via an optical cavity. A key question in such system is then, how to efficiently evaluate the photon reflection and transmission amplitudes, which are due to the medium's response corresponding to different pathways of scattering. In the past decades several approaches have been introduced to solve this problem. For example, one of the early approaches uses the Lippmann-Schwinger formalism in a Schr\"{o}dinger picture to evaluate the reflected and transmitted field amplitudes \cite{Shen07, Yud08, With10, Zheng10, Liao10}. This formulation, even though exact, cannot be applied for propagating photons interacting with separated multi-level emitters. Alternatively, some studies have used the transfer matrix method which is particularly useful in the weak excitation regime, where the emitters can be considered to be linear scatterers \cite{Deutsch95, Chang12}. 

To solve the problem of photon scattering from nonlinear emitters, an input-output formalism was developed although only for a two-level emitter coupled to a $1$D waveguide \cite{Fan10}. An analogous approach was later introduced for superconducting qubits coupled to a $1$D transmission line \cite{Lalu13}. There are several other frameworks to solve the scattering problem for nonlinear emitters coupled to $1$D waveguides \cite{Shi09, Shi11, Zheng11, Roy11, Laakso14}. Recently the formalism of Ref. \cite{Fan10} was generalized to multi-level emitters coupled to a $1$D waveguide \cite{Can15}. Furthermore, in a related work a path integral formalism-based scattering matrix was developed to study few-photon scattering dynamics in the non-Markovian regime \cite{Shi15}. Typically, all these approaches reduces to setting up the problem by either linearization, or by restricting the system to two-level emitters and a $1$D waveguide and then numerically solving it. Even then, the solution of the full photon-scattering problem from multi-level emitters in the paradigm of waveguide QED, remains quite tedious even for a single photon. 

In a preceding paper we developed a general photon-scattering relation from a system of multi-level quantum emitters embedded in a $3$-dimensional dielectric medium \cite{Das17}. The theoretical framework for this problem involved a set of excited and ground-state subspaces $M_{e}$ and $M_{g}$ respectively. Each of these subspaces are spanned by the manifold of the excited ($|e\rangle$) and ground ($|g\rangle$) states of the emitters. The theory is applicable to incident fields with a sufficiently low intensity, e.g., single-photon or weak coherent states, so that saturation effects can be ignored. In this limit, the coupling between the two subspaces can be treated perturbatively.  We showed that our theory provides a solution for the amplitudes of the scattered fields, in terms of the input-photon amplitude and the dynamical response of the emitters. As a continuation of Ref. \cite{Das17}, in this paper we apply the formalism to the particular case of $1$-dimensional waveguides and show how it can be used to solve a variety of scattering problems. Following Ref. \cite{Das17}, we derive a photon-scattering relation for a system of multi-level emitters coupled to a $1$D waveguide in the form 
\begin{figure*}[t!]
\includegraphics[height = 8.5 cm]{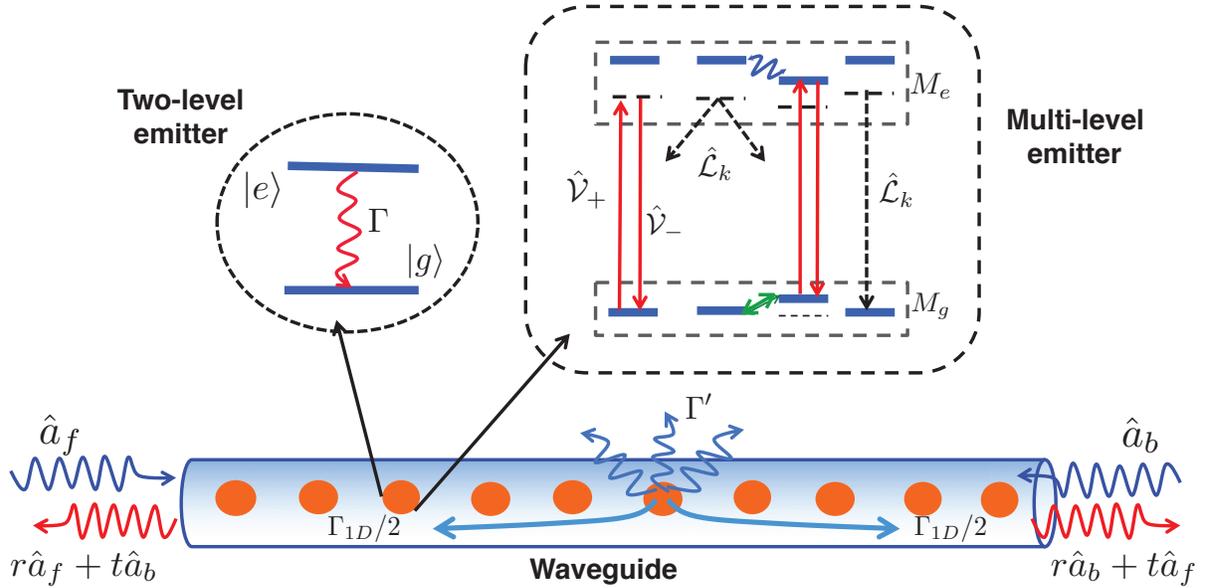}
\caption{Schematic of photon scattering from a generic system of emitters coupled to a waveguide. The emitters can be either a simple two-level system with a decay $\Gamma$ or have multiple levels. These can be separated into two subspaces: an excited-state manifold $M_e$ and a ground-state manifold $M_g$. The couplings between the two manifolds $\hat{\mathcal{V}}_{+}(\hat{\mathcal{V}}_{-})$ are assumed to be perturbative while the excited states experience decay modeled by the Lindblad operators $\hat{\mathcal{L}}_{k}$. The couplings within the excited and ground-state manifold are shown by the wiggly and straight arrow-headed lines respectively. The $1$D waveguide supports both forward and backward propagating modes of an input photon represented by the operators $a_f$ and $a_b$, respectively. Furthermore, the symbols $r$ and $t$ represent the reflection and transmission co-efficients satisfying the relation $|r|^{2}+|t|^{2} = 1$. Photons scattered from the emitters can decay to outside modes and into the waveguide with decay rates of $\Gamma'$ and $\Gamma_{1\text{D}}$, respectively.\label{figure1}}
\end{figure*}
\bea
\label{eqkey}
\hat{a}_{l,\text{out}} & = & \hat{a}_{l,\text{in}} + \sum_{l'}\sum_{gg'}\hat{\sigma}_{g'g}\mathcal{S}^{ll'}_{gg'}\hat{a}_{l',\text{in}}.          
\eea
Here $\hat{a}_{l,\text{in}}$ and $\hat{a}_{l,\text{out}}$ are the input and the output field-mode operators in the waveguide, $\hat{\sigma}_{g'g}$ is an operator in the Heisenberg picture giving the dynamics within the ground-state manifold $\{|g\rangle, |g'\rangle\}$ of the emitters, while the superscripts $(l,l')$ signify the directionality (forward, backward propagation) of photons in the waveguide. The kernel $\mathcal{S}^{ll'}_{gg'}$ is the scattering amplitude which can be evaluated once the coupling of the emitters has been determined.

In the following section we give a detailed derivation of Eq. (\ref{eqkey}) and discuss how to evaluate the ground-state dynamics in terms of the operator $\hat{\sigma}_{g'g}$. Furthermore, it will also be apparent that Eq. (\ref{eqkey}) has the following salient features $(a)$ it provides a direct solution of the scattering problem assuming Markovian dynamics for weak input fields, $(b)$ it can include any kind of dipole emitters coupled to the $1$D mode of a waveguide and $(c)$ it uses effective operators (EOs) to give a full solution of the emitter dynamics keeping track of all the phases and scattering component. The introduction of the EOs basically amounts to adiabatic elimination of the excited states and describing the system dynamics solely in terms of the ground-states evolution \cite{Reiter12}. Thus, by using EOs, the complications arising from multiple emitters in the scattering problem, can be reduced to solving the dynamics for the ground-state coherences and populations. 

The article is organized as follows: In Sec. II we give the detailed derivation of Eq. (\ref{eqkey}) starting from the photons scattering relation developed for a general dielectric medium in Ref. \cite{Das17}. In Sec. III we then elaborate on the physical processes that contribute to the non-Hermitian Hamiltonian, which is the key quantity for determining the scattering relation, and explain what the different terms in this Hamiltonian correspond to. Readers primarily interested in the application of the photon scattering formalism are encouraged to visit Sec. IV directly to avoid the technical details laid out in Secs. II and III. In Sec. IV we elaborate on our results by solving different examples of photon scattering from a single emitter coupled to a one-dimensional waveguide. We start with a simple example of a two-level emitter in Sec. IV.A and continue with a more complicated example of an emitter in a V-level configuration in Sec. IV.B. In Sec. IV.C we then consider several different cases of photon scattering from a system of multiple emitters coupled to a one-dimensional waveguide. In Sec. V we then give an example that demonstrates the versality of our formalism. We consider scattering from an emitter with multiple ground-states and study several aspects including the formation of ground-state superpositions conditioned on photon scattering. Finally, in Sec. VI we summarize our results and give an outlook. Several details of our calculations are relegated to the appendices. In Appendix A we provide the derivation of the photon-scattering relation for the $1$D waveguide. In Appendix B we present the derivation of the decay rate into the $1$D mode of the waveguide. In Appendix C we give details of the effective detunings and decays for the two-emitter systems.

\section{photon-scattering relation for emitters coupled to a one-dimensional waveguide}
In this section we derive the photon-scattering relation for a system of multi-level emitters coupled to a double-sided $1$D waveguide. To achieve this we first invoke the general photon-scattering relation in a dielectric medium 
\bea
\label{eqkey3}
\hat{\vec{\mathcal{E}}}^{+}(\vec{r},t) & = &\hat{\vec{\mathcal{E}}}_{in}(\vec{r},t)+\left(\frac{i\omega}{2\hbar}\right)\sum_{jj'}\sum_{gg'}\overleftrightarrow{\mathbf{G}}(\vec{r},\vec{r}_{j},\omega-\omega_{gg'})\nonumber\\
&\times&\hat{\sigma}_{g'g}\sum_{ee'}\bigg(\vec{d}^{j}_{ge}[\tilde{\mathcal{H}}_{nh}]^{-1}_{ee'}\vec{d}^{j'}_{e'g'}\bigg)\hat{\vec{\mathcal{E}}}_{in}(\vec{r}_{j'},t),
\eea
that was derived in Ref. \cite{Das17}. Here $\vec{r}$ is the point of observation, while $\vec{r}_{j}, \vec{r}_{j'}$ corresponds to the spatial positions of emitter $j$ and $j'$, respectively. The dipole moments $\vec{d}^{j}_{eg}$ and $\vec{d}^{j'}_{e'g'}$ correspond to the transition $|e\rangle \leftrightarrow |g\rangle$ and $|e'\rangle \leftrightarrow |g'\rangle$ for the emitters  $j$ and $j'$. The Green's function, $\overleftrightarrow{\mathbf{G}}(\vec{r},\vec{r}_{j},\omega-\omega_{gg'})$ gives the response of the field at the characteristic frequency $(\omega-\omega_{gg'})$ of the dielectric medium containing the emitters. Here $\omega$ is the central frequency of the input field and $\omega_{gg'} = (\omega_{g}-\omega_{g'})$ is the difference in frequency between states in the ground-state subspace. The input field in the above equation is defined as $\hat{\vec{\mathcal{E}}}_{in}(\vec{r},t) = i\sum_{k}\sqrt{\frac{\hbar\omega_{k}}{2}}\vec{F}_{k}(\vec{r})\hat{a}_{k}(0)e^{-i\omega_{k}t}$, where $\vec{F}_{k}(\vec{r})$ is the mode function while $\hat{a}_{k}$ is the mode operator for the $k^{th}$ mode of the field. The second term in Eq. (\ref{eqkey3}) represents the whole scattering event. It gives the scattered field including the dynamical response of the emitters. It is formulated in terms of the operator $\hat{\sigma}_{g'g} = |g\rangle\langle g'|$ and the non-Hermitian Hamiltonian $\tilde{\mathcal{H}}_{nh}$, which describes the dynamics in the excited-state subspace $M_e$. The non-Hermitian Hamiltonian is well known in the theory of Montecarlo wave-functions \cite{}. In Sec. III. we will describe in detail the meaning of this $\tilde{\mathcal{H}}_{nh}$ for our model. The states $|g\rangle$ and $|g'\rangle$ belong to the ground-state manifold $M_g$ of the emitters as shown in Fig. (\ref{figure1}). Note that our definition of the operator $\hat{\sigma}_{g'g}$ can be considered unconventional since the order is reversed. As we will see later, this definition gives us a simple relation to the density matrix $\rho_{g'g} = \langle\hat{\sigma}_{g'g}\rangle$ and simplifies the notation below. 

To proceed we first rewrite Eq. (\ref{eqkey}) in a more convenient form. We expand $\hat{\vec{\mathcal{E}}}_{in}(\vec{r},t)$ in terms of the Green's function
\bea
\label{eq1}
\hat{\vec{\mathcal{E}}}_{in}(\vec{r},t) = \int d\vec{r^\prime}~\epsilon(\vec{r^\prime})\overleftrightarrow{\mathbf{G}}(\vec{r},t,\vec{r^\prime},0)\hat{\vec{\mathcal{E}}}^{+}(\vec{r^\prime},0)
\eea
in Eq. (\ref{eqkey3}) and writing the frequency-dependent $\mathbf{G}(\vec{r},\vec{r}_{j},\omega-\omega_{gg'})$ as the Fourier transform of the time-dependent Green's function we get,
\bea
\label{eq2}
\hat{\vec{\mathcal{E}}}^{+}(\vec{r},t) & = &\int d\vec{r^{'}}~\epsilon(\vec{r^{'}})\overleftrightarrow{\mathbf{G}}(\vec{r},t,\vec{r^{'}},0)\hat{\vec{\mathcal{E}}}^{+}(\vec{r^{'}},0)+\left(\frac{i\omega}{2\hbar}\right)\nonumber\\
&\times&\sum_{gg'}\int^{t}_{-\infty} d\tau ~e^{i\omega_{gg'}(t-t')}\hat{\sigma}_{g'g}\overleftrightarrow{\mathbf{G}}(\vec{r},t,\vec{r_{j}},t')\nonumber\\
&\times&\sum_{ee'}\bigg(\vec{d}^{j}_{ge}[\tilde{\mathcal{H}}_{nh}]^{-1}_{ee'}\vec{d}^{j'}_{e'g}\bigg)\int d\vec{r^{'}}~\epsilon(\vec{r^{'}})\nonumber\\
&\times&\overleftrightarrow{\mathbf{G}}(\vec{r_{j'}},t',\vec{r^{'}},0)\hat{\vec{\mathcal{E}}}^{+}(\vec{r^{'}},0). 
\eea
Here $\epsilon(\vec{r})$ is the space-dependent electric permittivity of the waveguide. The first term on the right hand side of Eq. (\ref{eq2}) represents the freely propagating field with the Green's function being simply a propagator. 

We want to derive the photon-scattering relation for a double-sided $1$D waveguide. As such, we assume that the waveguide modes allow for the scattered photons to travel both in the forward $(f)$ and backward $(b)$ directions with wave-numbers $(k_f)$ and $(k_b)$, respectively. Furthermore, to account for the scattering into the waveguide and to the outside we divide $\hat{\mathcal{E}}^{+}_{k_{\zeta}}(\vec{r},t)$ into a waveguide and a radiative part. To treat this formally, we decompose the electric field in the form $\hat{\mathcal{E}}^{+}(\vec{r},t)  = \sum_{k_\zeta}\hat{\mathcal{E}}^{+}_{k_\zeta}(\vec{r},t)+\hat{\mathcal{E}}^{+}_{\text{rest}}(\vec{r},t)$ with $\zeta = \{f,b\}$, such that
\bea
\label{eq3}
\hat{\mathcal{E}}^{+}_{k_\zeta}(\vec{r},t)& = &i\sum_{k_{\zeta}}\sqrt{\frac{\hbar\omega_{k_{\zeta}}}{2}}\vec{F}_{k_{\zeta}}(r_{\perp})\hat{a}_{k_{\zeta}}e^{i(k_\zeta \text{z}-\omega t)},
\eea
represent the field in the forward and backward propagating modes of the waveguide. Here $i\sum_{k_{\zeta}}\sqrt{\frac{\hbar\omega_{k_{\zeta}}}{2}}\vec{F}_{k_{\zeta}}(r_{\perp})e^{ik_\zeta \text{z}}$ are the modes representing the field in the waveguide, $\text{z}$ is the co-ordinate along the waveguide, while $\mathcal{E}^{+}_{\text{rest}}(\vec{r},t)$ are the radiative modes representing the scattered light to the outside. 

Substituting Eq. (\ref{eq3}) into Eq. (\ref{eq2}) and decomposing the Green's function into the forward, backward and the rest of the components as
\bea
\label{eq4} 
\overleftrightarrow{\mathbf{G}}(\vec{r},t,\vec{r'},t') &= &\sum_{\zeta}\overleftrightarrow{\mathbf{G}}_{\zeta}(\vec{r}_{\perp},t,\vec{r'}_{\perp},t')+\overleftrightarrow{\mathbf{G}}_{\text{rest}}(\vec{r},t,\vec{r'},t'),\nonumber\\
\eea
we arrive finally (see Appendix A for details) at the photon-scattering relation in the $1$D waveguide 
\bea
\label{eq5}
\hat{a}_{\zeta, \text{o}}(\text{z},t) & = & \hat{a}_{\zeta, \text{in}}(\text{z}\mp v_{g}t)+i\sum_{\zeta'}\sum_{gg'}\hat{\sigma}_{g'g}[\mathcal{S}^{\zeta\zeta'}_{gg'}]_{\mp}\nonumber\\
&\times& \hat{a}_{\zeta', \text{in}}(\text{z}\mp v_{g}t)+\mathcal{F}.
\eea
Here $v_{g}$ is group velocity of the photon in the waveguide, while $\mathcal{F}$ is a noise operator that corresponds to the $\mathbf{G}_{\text{rest},\zeta}$ and $\mathcal{E}^{+}_{\text{rest}}$ and is associated with the loss of photons out of the waveguide. The mode operators $\hat{a}_{\text{o}}$ and $\hat{a}_{\text{in}}$ correspond to the output and input light field, respectively. Note that the $``-" (+)$ sign stands for photons travelling in the forward (backward) direction. The scattering amplitude $[\mathcal{S}^{\zeta\zeta'}_{gg'}]_{\mp}$ is defined as 
\bea
\label{eq6}
[\mathcal{S}^{\zeta\zeta'}_{gg'}]_{\mp}& = &\sum_{jj'}\sum_{ee'} \mathcal{A}^{\dagger j \zeta}_{ge(1\text{D})}[\tilde{\mathcal{H}}_{nh}]^{-1}_{ee'}\mathcal{A}^{j'\zeta'}_{e'g'(1\text{D})}\nonumber\\
&\exp&[\mp i((k_\zeta-k_{\zeta'})\text{z}_{j}+\omega_{g'g}(\text{z}-\text{z}_{j})/v_{g}],
\eea
where we have defined the directional coupling of the emitters to the waveguide mode as 
\bea
\label{eq6aa}
\mathcal{A}^{j\zeta}_{eg} =  \sqrt{\frac{\pi\omega}{\hbar v_{g}}}\left[\vec{d}^{j}_{eg}\cdot\vec{F}_{k_\zeta}(r_{j_\perp})\right]. 
\eea
with $\mathcal{A}^{\dagger j \zeta}_{ge(1\text{D})} = \mathcal{A}^{\ast j \zeta}_{eg(1\text{D})}$. The wave vectors in the forward and backward direction follow the relation $\Delta{k} = (k_\zeta-k_{\zeta'}) =  0~\text{and}~2{k_{0}}$ for $\zeta = \zeta'$ and $\zeta \neq \zeta'$, respectively. The photon-scattering relation in Eq. (\ref{eq5}) is the key result of this work and has the generic form stated in Eq. (\ref{eqkey}). Note that, the coupling defined in Eq. (\ref{eq6aa}) has a directional dependence and in principle its strength can be different for the field-mode propagating along two different directions (forward or backward) in the waveguide. This leads to an interesting and emerging question of chiral light-matter interaction \cite{Peter17}. Even though we do not explicitly address this, our general formalism is already equipped with such possibilities. As such the photon-scattering relation in Eq. (\ref{eq5}) is applicable even to the study of  chiral interactions in waveguides. 

It is worth emphasizing that in the derived photon-scattering relation all the system properties are included through the non-Hermitian Hamiltonian $\tilde{\mathcal{H}}_{nh}$ while the evolution of the emitters, response is through the operator $\hat{\sigma}_{g'g}$ defined in the ground-state manifold $M_g$. To get the complete photon-scattering dynamics using the photon-scattering relation introduced above we need to find $\hat{\sigma}_{g'g}$. This can be quite cumbersome for complex systems involving multiple levels. However, by exploiting the formulation of EOs \cite{Reiter12}, which again involves the inverse of the non-Hermitian Hamiltonian $[\tilde{\mathcal{H}}_{nh}]$, we can solve for $\hat{\sigma}_{g'g}$ using the master equation derived explicitly in the preceding paper \cite{Das17} 
\bea
\label{eq6a}
\dot{\hat{\sigma}} & = &:~ i\left[\hat{\mathcal{H}}_{\text{eff}}, \hat{\sigma}\right]-\frac{1}{2}\sum_{k}\left(\hat{\mathcal{L}}^{k\dagger}_{\text{eff}}\hat{\mathcal{L}}^{k}_{\text{eff}}\hat{\sigma}+\hat{\sigma}\hat{\mathcal{L}}^{k\dagger}_{\text{eff}}\hat{\mathcal{L}}^{k}_{\text{eff}}\right)\nonumber\\
&+&\sum_{k}\hat{\mathcal{L}}^{k}_{\text{eff}}\hat{\sigma}\hat{\mathcal{L}}^{k\dagger}_{\text{eff}}~:.
\eea
Here all the operators are defined in the Heisenberg picture and the subscript ``eff" symbolizes EO's. The symbol $``~:~"$ in Eq. (\ref{eq6a}) stands for normal ordering, the significance of which will be discussed in details in section VI.C. Note that, Eq. (\ref{eq6a}) is a Heisenberg-picture generalization of the result of Ref. \cite{Reiter12} to quantum fields. Solving the above master equation for a given system is a straightforward algebraic/numerical exercise whose complexity simply depends on the size of the Hilbert space of the emitters. Later in section IV.C we consider an example where the emitters have multiple ground-states and show how one can use the master equation in Eq. (\ref{eq6a}) to solve for the dynamics of the emitter's ground-state. 

It is important to point out that for the examples we discuss in Sec. IV, the noise term $\mathcal{F}$ in Eq. (\ref{eq5}) is typically neglected. This is justified by the fact that in those examples we are only interested in the click probability where the vacuum noise does not contribute to any photodetector clicks. However, we would like to remind the readers that in general particular care should be taken for Heisenberg equations as the noise can play a crucial role in the system dynamics. We account for this in our formalism through the effective Lindblad operators in the master equation, which includes the noise contribution. Hence for problems where the scattering is influenced by the coherence dynamics of the ground-states, the crucial effect of noise is taken care of in the master equation. We show this in detail in the example in Sec. VI.C. Thus we discuss explicitly how to deal with the noise and treat it via the effective-operator master equation.
\section{The non-Hermitian Hamiltonian}
To be able to apply our formalism, it is important to understand the non-Hermitian Hamiltonian $[\tilde{\mathcal{H}}_{nh}]$ in Eq. (\ref{eq6}). The general form of the non-Hermitian Hamiltonian from \cite{Das17} is
\bea
\label{eq7}
\left[\mathcal{H}_{nh}\right]_{ee'} = \left[\mathcal{H}_{c_{e}}\right]_{ee'}-i\sum_{jj'}\sum_{g}\left(\frac{1}{2}\Gamma^{jj' e'e}_{gg}-i\Omega^{jj'e' e}_{gg}\right).\nonumber\\
\eea
Note that this non-Hermitian Hamiltonian includes all possible interactions that the emitters can have within the excited-state manifold. In the following we discuss each of the terms in Eq. (\ref{eq7}). The first term $\mathcal{H}_{c_{e}}$ is the Hamiltonian of the system defined in the single excitation manifold $M_e$ as shown in Fig. \ref{figure1}. Note that this term is completely general and can in principle also include effects like the long-range Rydberg interactions among emitters. The second and third term $\Gamma^{jj' e'e}_{gg}$ and $\Omega^{jj'e' e}_{gg}$ arise from the dynamics induced by the quantized field and are related to the decay from the manifold $M_e$ to $M_g$, and shifts of the states in the manifold $M_e$ due to light induced coupling between the emitters. They are defined as
\bea
\label{eq8a}
\Gamma^{jj',e'e}_{gg} & = &\frac{2\omega_{eg'}^{2}}{\hbar c^2}\left\{\vec{d}^{j}_{e'g'}\cdot\mathbf{Im}\overleftrightarrow{\mathbf{G}}(\vec{r}_{j},\vec{r}_{j'},\omega_{e'g})\cdot\vec{d}^{j'}_{ge}\right\},\nonumber\\
\\
\label{eq8b}
\Omega^{jj',e'e}_{gg} & = &\mathbf{P}\int d\omega \left(\frac{\omega^2}{\hbar\pi c^2}\right)\bigg\{\frac{\vec{d}^{j}_{e'g}\cdot\mathbf{Im}\overleftrightarrow{\mathbf{G}}\cdot\vec{d}^{j'}_{ge}}{(\omega-\omega_{e'g'}+i\epsilon)}\bigg\},\nonumber\\
\eea
where the excited $|e\rangle$, $|e'\rangle$ and ground $|g\rangle$ states belong to the excited and ground subspaces $M_e$ and $M_g$, respectively. Note that to write Eq. (\ref{eq8a}) and Eq. (\ref{eq8b}) we have used the general 
form of these expression derived in Ref. \cite{Das17}.

The $\textbf{Im}\overleftrightarrow{G}$ in the above set of equations stands for imaginary part of the Green's tensor. On expanding the Green's function using Eq. (\ref{eq4})
and substituting it in Eqs. (\ref{eq8a}) and (\ref{eq8b}) we get,
\bea
\label{eq9a}
\Gamma^{jj',e'e}_{gg} & = &\frac{2\omega_{e'g'}^{2}}{\hbar v_{g}^2}\left\{\vec{d}^{j}_{e'g}\cdot\sum_{\zeta}\mathbf{Im}\overleftrightarrow{\mathbf{G}}_\zeta(\vec{r}_{j},\vec{r}_{j'},\omega_{e'g})\cdot\vec{d}^{j'}_{ge}\right\},\nonumber\\
&+&\frac{2\omega_{e'g}^{2}}{\hbar c^2}\left\{\vec{d}^{j}_{e'g}\cdot\mathbf{Im}\overleftrightarrow{\mathbf{G}}_{\text{res}t}(\vec{r}_{j},\vec{r}_{j'},\omega_{e'g})\cdot\vec{d}^{j'}_{ge}\right\},\nonumber\\
\\ 
\label{eq9b}
\Omega^{jj',e'e}_{gg} & = &\mathbf{P}\int d\omega \left(\frac{\omega^2}{\hbar\pi v_{g}^2}\right)\bigg\{\frac{\vec{d}^{j}_{e'g}\cdot\sum_{\zeta}\mathbf{Im}\overleftrightarrow{\mathbf{G}}_\zeta\cdot\vec{d}^{j'}_{ge}}{(\omega-\omega_{e'g}+i\epsilon)}\bigg\},\nonumber\\
\\
&+&\mathbf{P}\int d\omega \left(\frac{\omega^2}{\hbar\pi c^2}\right)\bigg\{\frac{\vec{d}^{j}_{e'g}\cdot\mathbf{Im}\overleftrightarrow{\mathbf{G}}_{\text{rest}}\cdot\vec{d}^{j'}_{ge}}{(\omega-\omega_{e'g}+i\epsilon)}\bigg\}\nonumber\\
\eea
We rewrite $\Gamma^{jj',e'e}_{gg}$ in Eq. (\ref{eq9a}) in the form $\Gamma^{jj',e'e}_{gg}  = [\Gamma^{jj',e'e}_{gg}]_{\text{w}} +[\Gamma^{jj',e'e}_{gg}]_{\text{rest}}$. Here $[\Gamma^{jj',e'e}_{gg}]_{\text{w}}$ corresponds to the first term on the right-hand side of Eq. (\ref{eq9a}) and represents decay-induced coupling between the emitters mediated by the $1$D waveguide mode. $[\Gamma^{jj',e'e}_{gg}]_{\text{rest}}$ represents the second term and arises due to collective decay to the non-waveguide modes (decay to the outside of the waveguide).  For $j = j'$, $[\Gamma^{jj',e'e}_{gg}]_{\text{w}}$ corresponds to spontaneous decay of the emitter into the $1$D waveguide mode while $[\Gamma^{jj',e'e}_{gg}]_{\text{rest}}$ gives spontaneous decay of the emitter to the outside of the waveguide. Similarly, Eq. (\ref{eq9b}) for $j \neq j'$ can be defined as $\Omega^{jj',e'e}_{gg} = [\Omega^{jj',e'e}_{gg}]_{\text{w}}+[\Omega^{jj',e'e}_{gg}]_{\text{rest}}$, where $[\Omega^{jj',e'e}_{gg}]_{\text{w}}$ represent the first term on the right-hand side of Eq. (\ref{eq9b}) and stands for waveguide-mediated coupling of the emitters while $[\Omega^{jj',e'e}_{gg}]_{\text{rest}}$ represents the second term and corresponds to coupling via other processes like dipole-dipole interactions. For $j = j'$, the coupling $\Omega^{jj',e'e}_{gg}$ gives a contribution to the Lamb shift of the excited state of a single emitter. Note that in Ref. \cite{Das17} these terms were derived within the rotating wave approximation, which does not produce the correct form of the dipole-dipole interaction for emitters separated by less than a wavelength. Care should therefore be taken to use the correct shifts beyond the rotating wave approximation for nearby emitters. 

In the following we derive an exact expression for the waveguide-mediated coupling between the emitters, by solving for the first terms on the right-hand side of Eq. (\ref{eq9a}) and Eq. (\ref{eq9b}). For this purpose we invoke the relation \cite{Novobook06}
\bea
\label{eq9}
\sum_{k}\omega_k \vec{F}_k(\vec{r})\vec{F}^\ast_k(\vec{r'})~e^{-i\omega_k(t-t')} & = & 2\int~d\omega~e^{-i\omega(t-t')} \frac{\omega^2}{\pi c^2}\nonumber\\
&\times&\textbf{Im}\{\overleftrightarrow{G}(\vec{r},\vec{r'},\omega)\},
\eea                                             
and do an inverse Fourier transform of it to get
\bea
\label{eq10}
\mathbf{Im}\overleftrightarrow{\mathbf{G}}_\zeta (\vec{r}_j,\vec{r}_{j'}, \omega) & = &\frac{\pi}{k_\zeta}F_{k_\zeta}(\vec{r}_{j\perp})F^\ast_{k_\zeta}(\vec{r}_{j'\perp})\nonumber\\
&\times&\cos\left(k_\zeta|\text{z}_j-\text{z}_{j'}|\right),
\eea
where $k_\zeta = \pm~\omega/v_g$, with the $+(-)$ sign corresponding to the forward (backward) propagation direction. Then substituting Eq. (\ref{eq10}) into the first term on the right-hand side of Eq. (\ref{eq9a}) and on using Eq. (\ref{eq6aa}) we get 
\bea
\label{eq11}
[\Gamma^{jj',e'e}_{gg}]_{\text{w}} = 2\sum_\zeta\mathcal{A}^{ j \zeta}_{e'g(1\text{D})}\mathcal{A}^{\dagger j'\zeta}_{ge(1\text{D})}\cos\left(k_\zeta|\text{z}_j-\text{z}_{j'}|\right).\nonumber\\
\eea
Furthermore, substituting Eq. (\ref{eq10}) into the first term on the right-hand side of Eq. (\ref{eq9b}) and then performing the principal value integral over an anticlockwise contour and 
invoking Cauchy's residue theorem (see Appendix B for details) gives us 
\bea
\label{eq12}
[\Omega^{jj',e'e}_{gg}]_{\text{w}} = -\sum_\zeta\mathcal{A}^{ j \zeta}_{e'g(1\text{D})}\mathcal{A}^{\dagger j'\zeta}_{ge(1\text{D})}\sin\left(k_\zeta|\text{z}_j-\text{z}_{j'}|\right).\nonumber\\
\eea

If we refer to the expression for the non-Hermitian Hamiltonian in Eq. (\ref{eq7}) and consider the contribution to the second and the third term due to the waveguide-mediated interactions, we find, using Eq. (\ref{eq11}) and Eq. (\ref{eq12}), that  \cite{Chang12, Can15, Hak05}
\bea
\label{eq13}
\frac{1}{2}[\Gamma^{jj',e'e}_{gg}]_{\text{w}} - i[\Omega^{jj',e'e}_{gg}]_{\text{w}} & = &\sum_\zeta\mathcal{A}^{ j \zeta}_{e'g(1\text{D})}\mathcal{A}^{\dagger j'\zeta}_{ge(1\text{D})}\nonumber\\
&\times&e^{ik_\zeta|\text{z}_j-\text{z}_{j'}|}.
\eea
Note that for the case of a single two-level emitter, $j=j'$ and $e' = e$. Eq. (\ref{eq13}) becomes 
\bea
\label{eq14}
\sum_{\zeta}|\mathcal{A}^{\zeta}_{eg(1\text{D})}|^2 = \sum_{\zeta}\Gamma^{e,\zeta}_{g,1\text{D}} = \Gamma^{e}_{g,1\text{D}},
\eea
where $\Gamma^{e}_{g,1\text{D}}$ is the total decay of energy level $|e\rangle$ into the $1$D mode of the waveguide for the emitter transition $|e\rangle\rightarrow|g'\rangle$ . 

We can now rewrite the non-Hermitian Hamiltonian $[\mathcal{H}_{nh}]_{ee'}$ of Eq. (\ref{eq7}) as a combination of two parts, one comprising of all the interactions mediated by the waveguide (w) while the other one concerning all other processes not mediated by the waveguide (nw). The non-Hermitian Hamiltonian then takes the form $[\tilde{\mathcal{H}}_{nh}]_{ee'} = \left[\tilde{\mathcal{H}}_{c}\right]_{ee'}+\left[(\tilde{\mathcal{H}}_{nh})_{\text{w}}\right] _{ee'}+\left[(\tilde{\mathcal{H}}_{nh})_{\text{nw}}\right] _{ee'}$, where 
\bea
\label{eq15}
\left[(\tilde{\mathcal{H}}_{nh})_{\text{w}}\right] _{ee'} & = &-i\sum_{jj'}\sum_{g,\zeta}\mathcal{A}^{ j \zeta}_{e'g(1\text{D})}\mathcal{A}^{\dagger j'\zeta}_{ge(1\text{D})}e^{ik_\zeta|\text{z}_j-\text{z}_{j'}|}\nonumber\\
\\
\label{eq16}
\left[(\tilde{\mathcal{H}}_{nh})_{\text{nw}}\right] _{ee'} &= &-\sum_{jj'}\sum_{g}\left(\frac{i}{2}[\Gamma^{jj',e'e}_{gg}]_{\text{rest}} +[\Omega^{jj',e'e}_{gg}]_{\text{rest}} \right).\nonumber\\
\eea
Here $\tilde{\mathcal{H}}_{c} = \mathcal{H}_{c_e}-E_g-\hbar\omega$, with $E_g$ being the energy of the ground-state involved in the excitation process while $\omega$ is the frequency of the incoming photon.
The waveguide-mediated off-diagonal term in Eq. (\ref{eq15}) can also be re-written in terms of $\Gamma_{1\text{D}}$ as,
\bea
\label{eq18}
\left[(\tilde{\mathcal{H}}_{nh})_{\text{w}}\right] _{ee'} & = &-i\sum_{jj'}\sum_{g\zeta}\sqrt{\Gamma^{e'j\zeta}_{g,1\text{D}}}\sqrt{\Gamma^{ej'\zeta}_{g,1\text{D}}}e^{i\left(\phi_{e'g'}-\phi_{eg} \right)}\nonumber\\
&\times&e^{ik_\zeta|\text{z}_j-\text{z}_{j'}|},
 \eea
where we have used $\mathcal{A}^{ j \zeta}_{eg(1\text{D})} = |\mathcal{A}^{ j \zeta}_{eg(1\text{D})}|e^{i\phi_{eg}}$ and the definition of directional decay into the waveguide $\Gamma^{e\zeta}_{g,1\text{D}}$ in terms of the coupling constants $\mathcal{A}$ from Eq. (\ref{eq14}). 

On using the general form of $[\tilde{\mathcal{H}}_{nh}]_{ee'} $ and Eq. (\ref{eq18}) we find that the non-Hermitian Hamiltonian has a simple diagonal part $(j = j')$ spanned by the excited states of the emitters as
\bea
\label{eq17}
\left[\tilde{\mathcal{H}}_{nh}\right]_{ee} &= & \tilde{\Delta}_e-\frac{i}{2}\Gamma_e,
\eea
where $\tilde{\Delta}_e = [\tilde{\mathcal{H}}_{c_e}-E_g-\hbar\omega]_{ee}$ and $\Gamma_{e} = \Gamma_{e}' + \Gamma_{e(1\text{D})} = \sum_{g}\left[\Gamma^{e}_{g \text{rest}}+\sum_{\zeta}\Gamma^{e,\zeta}_{g,1\text{D}}\right],$ is the natural line width of an excited state $|e\rangle$ in the single-excitation manifold $M_e$. Here $\Gamma_{e}' = \sum_{g}\Gamma^{e}_{g\text{rest}}$ is the total decay rate to the outside of the waveguide and $\tilde{\mathcal{H}}_{c_e}$ is a redefined excited-state Hamiltonian formed by absorbing the Lamb-shift contribution in $\mathcal{H}_{c_e}$. Note that Eq. (\ref{eq17}) can also be written in the standard form of a non-Hermitian Hamiltonian 
\bea
\label{eq17a}
\hat{\mathcal{H}}_{nh} & = & \hat{\tilde{\mathcal{H}}}_{c_e}-\frac{i}{2}\sum_k\hat{\mathcal{L}}^\dagger_k \hat{\mathcal{L}}_k, 
\eea
where the Lindblad operators $\hat{\mathcal{L}}_k$ model decay of an excited emitter both into and outside of the waveguide. 

We next discuss the contribution to the non-Hermitian Hamiltonian from the non-waveguide part $(\tilde{\mathcal{H}}_{nh})_{\text{nw}}$ in Eq. (\ref{eq16}). These terms can have contributions both for inter- and intra-emitter couplings. In the Dicke superradiant limit, where the separation between the emitters is less than a wavelength, the $(\tilde{\mathcal{H}}_{nh})_{\text{nw}}$ gives rise to collective decay and dipole-dipole couplings. For most of this article we will ignore the $(\tilde{\mathcal{H}}_{nh})_{\text{nw}}$ part of the non-Hermitian Hamiltonian. However, we do use this in two particular examples to illustrate the wide range of applicability of our formalism.
\section{Application of the formalism to emitters with a single ground-state}
In the previous sections we have introduced a formalism for photon scattering from quantum emitters in a $1$D waveguide, and elaborated on the non-Hermitian Hamiltonian that is central to the response of the emitters interacting with the incoming field. In the following sub-sections IV.A - IV.C we focus on, a number of paradigmatic physical situations that demonstrates the effectiveness of our formalism for solving photon scattering problems in waveguides. In this section we restrict ourselves to examples where the emitters have a single ground-state. In the next section we consider in detail an example of emitters with multiple ground-states. It is worth emphasizing that even the simple and generic examples of scattering that we treat here are in some cases rather tedious to solve with the existing methods. However, using our formalism we can immediately provide the solution to these problems. Note that for notational convenience, in all further discussion we will label the photons incoming from the left and moving to the right with subscript (R) and the photons moving to the left as (L), such that now $\zeta = \{\text{R, L}\}$.
\subsection{A two-level emitter coupled to a one-dimensional waveguide}
\begin{figure}
	\includegraphics[width = 0.45\textwidth]{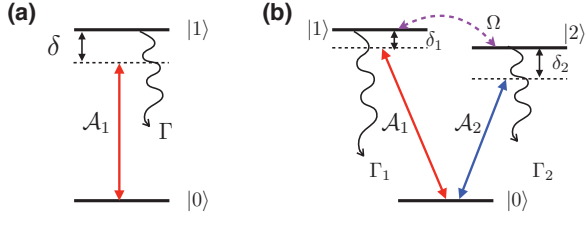} 
	\caption{Schematic diagram of the energy level structure of emitters with (a) single optical transition (b) two optical transitions in V-configuration. Here $|0\rangle$ is the ground-stateand $|i = 1,2\rangle$ the excited states of the emitter. The linewidth of the excited states is given by $\Gamma$'s and the $\delta$'s are detuning of the transition with respect to the frequency of the incoming photon. The coupling strength of the transitions to the waveguide mode is given by $\mathcal{A}$'s.}
	\label{fig2}
\end{figure}

We first analyze the simplest possible system. We consider an emitter comprising two levels with a single optical transition between a ground level $\ket{0}$ and an excited level $\ket{1}$ as shown schematically in Fig. \ref{fig2}~(a). The emitter is located at a position $\text{z}_0$ along the axis of a $1$D waveguide. The transition is coherently coupled to a waveguide. Such a system is generally described by a Hamiltonian $\hat{\mathcal{H}} = \hat{\mathcal{H}}_0 + \hat{\mathcal{V}}~(\hbar = 1)$, where
\begin{align}
\label{eq19}
\hat{\mathcal{H}}_0 &= \omega_{11}\hat{\sigma}_{11} + \omega_{00}\hat{\sigma}_{00}+ \hat{\mathcal{H}}_{F}.
\\
\hat{\mathcal{V}} &=\hat{\mathcal{V}}_{-} +\hat{\mathcal{V}}_{+},\nonumber\\
& = \sum_{\mu}\mathcal{A}^{\mu}_{10}\hat{a}_{\mu}^{\dagger}\hat{\sigma}_{10} +\sum_{\mu}\mathcal{A}^{\ast\mu}_{01}\hat{a}_{\mu}\hat{\sigma}_{01}
\end{align}
with the free-energy Hamiltonian $\hat{\mathcal{H}}_0$, and the Hamiltonian of the field being given by $\hat{\mathcal{H}}_{F}$, while the excitation (de-excitation) is represented by                                                                                                                                         $\hat{\mathcal{V}}_{+}$ ($\hat{\mathcal{V}}_{-} = [\hat{\mathcal{V}}_{+}]^\dagger)$. Here, $\omega_{11}$ and $\omega_{00}$ are the energies of levels $\ket{1}$ and $\ket{0}$, respectively. Furthermore, as above we have used the definition of the atomic operator $\hat{\sigma}_{ij} = \ket{j}\bra{i}$ such that the density matrix is given by $\rho_{ij} = \langle \hat{\sigma}_{ij}\rangle$. The coupling strength of the emitter transition $|i\rangle \leftrightarrow |j\rangle$ to the field is given by $\mathcal{A}^{\mu}_{ij}$,  with $a_{\mu}~(a_{\mu}^\dagger)$ being the corresponding annihilation (creation) field-mode operator and $\mu = \{\zeta, s\}$. Here, $(\mu = s)$ signifies that $\mathcal{A}^{s}_{ij}$ is the coupling strength of the transition to modes outside the waveguide, while $(\mu = \zeta)$ represents the directional coupling to the $1$D waveguide mode with strength $\mathcal{A}^{\zeta}_{ij,(1\text{D})}$. For the rest of this example we drop the subscripts $(i,j)$ from the coupling constants as it involves only a single transition. We can then write the non-Hermitian Hamiltonian for this system in the form $\hat{\mathcal{H}}_{nh} = \hat{\mathcal{H}}_{0}-\frac{i}{2}\sum_{k}\hat{\mathcal{L}}^\dagger_{k}\hat{\mathcal{L}}_{k}$, where the Lindblad operators $\mathcal{L}_{k}$ are given by
\bea
\label{eq19a}
\hat{\mathcal{L}}_{s} & = &\mathcal{A}^{s}~\hat{\sigma}_{10} = \sqrt{\Gamma'}~\hat{\sigma}_{10} ,\\
\hat{\mathcal{L}}_{\zeta_{(1\text{D})}}& = & \mathcal{A}^{\zeta}_{1\text{D}}~\hat{\sigma}_{10}= \sqrt{\Gamma^{\zeta}_{1\text{D}}}~\hat{\sigma}_{10},
\eea
corresponding to decay out of $(s)$ and into the waveguide $(\zeta)$. Note that in writing Eq. (\ref{eq19a}) we have used the definition of $\Gamma^\zeta_{1\text{D}}$ from Eq. (\ref{eq14}), and defined the rate of decay out of the waveguide as $\Gamma' = |\mathcal{A}^{s}|^{2}$. The non-Hermitian Hamiltonian can then be written similar to that in Eq. (\ref{eq17}) as
\begin{align}
\label{eq20}
\hat{\mathcal{H}}_{nh} = \left(\delta - \frac{i \Gamma}{2}\right) \hat{\sigma}_{11} \equiv \tilde{\delta} \hat{\sigma}_{11},
\end{align}
where $\Gamma$ is the total decay rate of the level $\ket{1}$ into $\ket{0}$ and is given by $\Gamma = \Gamma'+\sum_\zeta\Gamma^\zeta_{1\text{D}}$, while the detuning is $\delta = \omega_{11} - \omega_{00} - \omega$. Here $\omega$ is the frequency of the incoming field. Combining the decay with the detuning we then define $\tilde{\delta} = (\delta - i\Gamma/2)$ as the complex energy of the state $\ket{1}$. Inverting the $\langle 1|\hat{\mathcal{H}}_{nh}|1\rangle$ is then straightforward and we find
\begin{align}
\label{eq21}
\hat{\mathcal{H}}_{nh}^{-1} = \tilde{\delta}^{-1} \hat{\sigma}_{11},
\end{align}

For a single photon incident from left and propagating towards the right in the waveguide, Eq. (\ref{eq5}) straightway gives the complete scattering dynamics of the photon from the two-level emitter. Let us write Eq. (\ref{eq5}) in terms of the field-mode operators on the left and right of the emitter, after scattering of a photon as
\bea
\label{eq22}
\hat{a}_\text{out,R}(\text{z},t) & = & \left[1+i\Gamma^\text{R}_{1\text{D}}\tilde{\delta}^{-1}\hat{\sigma}_{00}\right]\hat{a}_\text{in,R}(\text{z}-v_gt),\nonumber\\
\\
\label{eq23}
\hat{a}_\text{out,L}(\text{z}',t) & = & i\left[\sqrt{\Gamma^{\text{L}}_{1\text{D}}}\left(\tilde{\delta}^{-1}\right)\sqrt{\Gamma^{\text{R}}_{1\text{D}}}~\right]e^{2ik_{0}(\text{z}_{0}-\text{z}')}\nonumber\\
&\times&\hat{\sigma}_{00}\hat{a}_\text{in,R}(\text{z}'+v_g t),
\eea
where we have used that $(\hat{\mathcal{H}}_{nh})^{-1}_{11} = \tilde{\delta}^{-1}$ and $\text{z} (\text{z}')$ is the point of observation to the right (left) of the emitter spatially situated at $\text{z}_{0}$. Here $e^{2ik_{0}(\text{z}_{0}-\text{z}')}$ is an additional phase that the reflected photon picks up as it propagates towards the left of the emitter. Note that in writing Eq. (\ref{eq22}) and Eq. (\ref{eq23}) we have neglected the noise term as we are mainly concerned with the photon click probability at a detector.  

\begin{figure}
\begin{tabular}{ccc}
	\includegraphics[width = 0.225\textwidth]{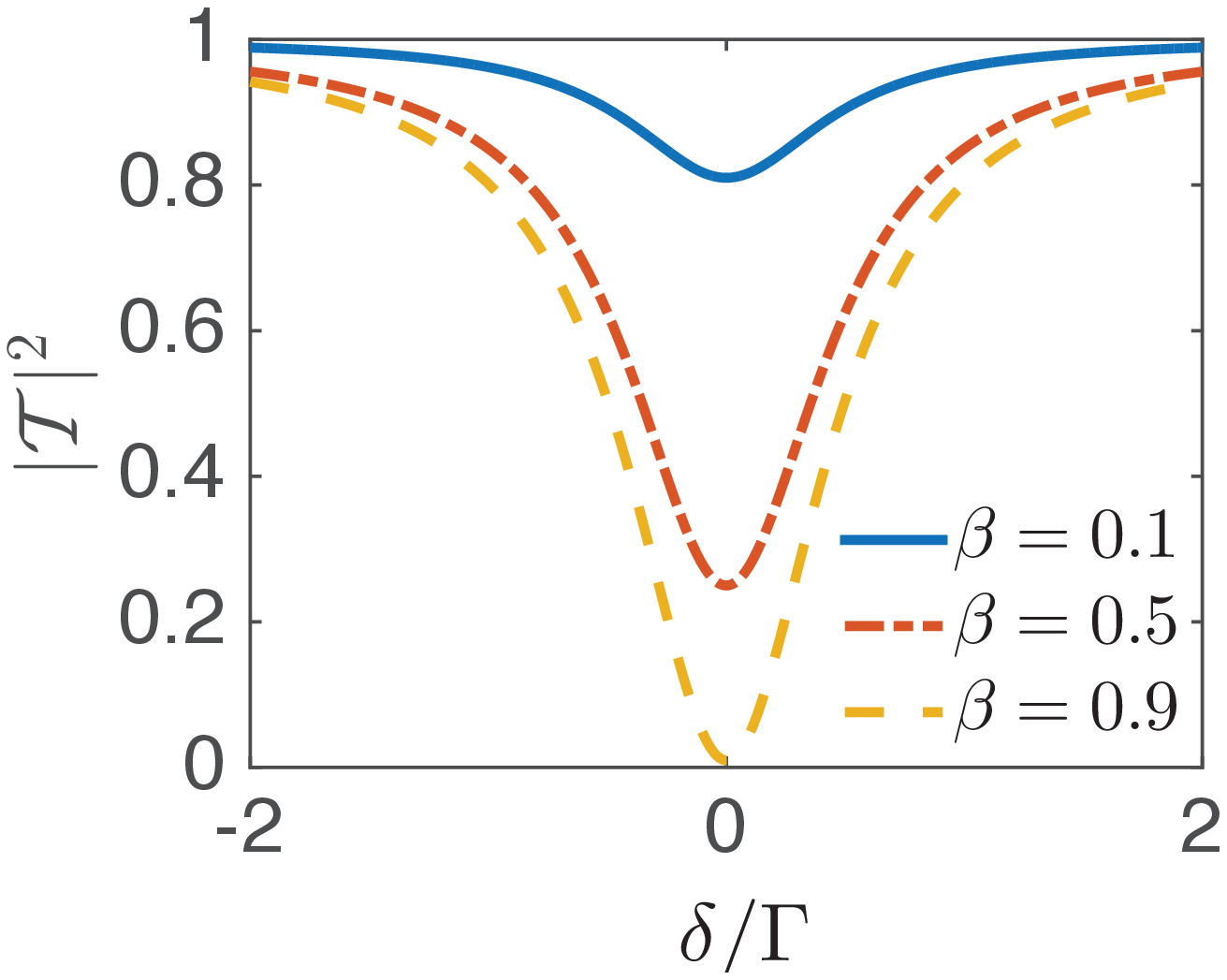} & \includegraphics[width = 0.265\textwidth]{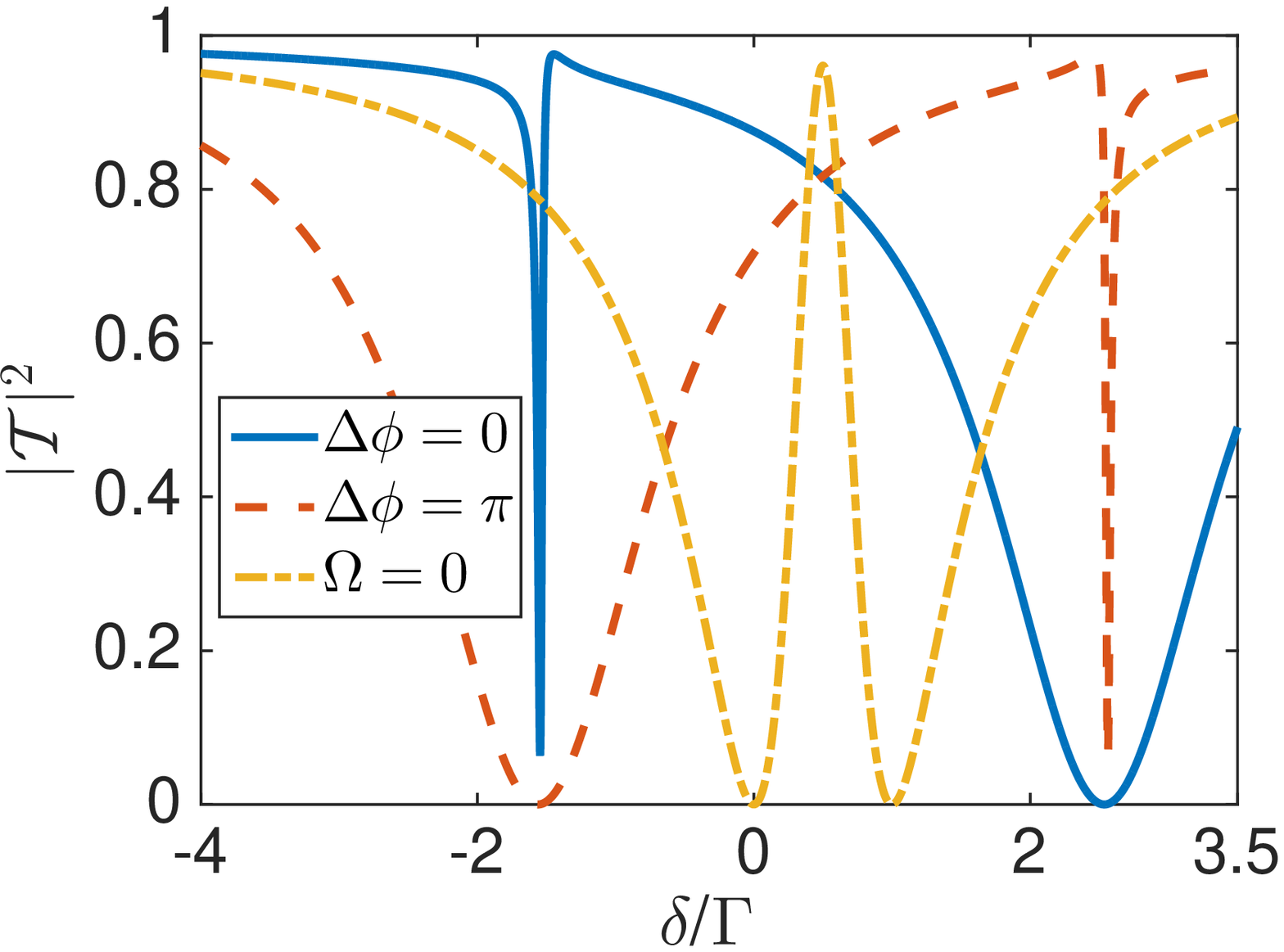}\\
         (a) & (b)
\end{tabular}	
	\caption{Transmitted intensity $|\mathcal{T}|^2=|\langle \hat{a}^\dagger_{out,\text{R}} \hat{a}_{out,\text{R}}\rangle/\langle \hat{a}^\dagger_{in} \hat{a}_{in}\rangle|$ for a single (a) two-level emitter and, (b) three-level emitter in the V-configuration coupled to a 1\text{D} waveguide. For (a) we consider the parameters, $\delta=\omega_{11}-\omega_{00}-\omega$ and different values of $\beta$ while for (b) we consider $\delta_1=-\delta$, $\delta_2=\Gamma-\delta$, $\beta=0.99$, coupling $\Omega=2\Gamma$ or $0$, and we plot the results for $\Delta\phi = 0$ and $\Delta\phi = \pi$.}
	\label{Spec1}
\end{figure}

Substituting for $\tilde{\delta}$ and assuming that $\Gamma^{\text{R}}_{1\text{D}} = \Gamma^{\text{L}} _{1\text{D}} = \Gamma_{1\text{D}}/2$, we get  
\bea
\label{eq24}
\hat{a}_\text{out,R}(\text{z},t) & = & \left[1-\frac{\Gamma_{1\text{D}}}{\Gamma+2i\delta}\right]\hat{a}_\text{in,R}(\text{z}-v_g t),\\
\label{eq25}
\hat{a}_\text{out,L}(\text{z}',t) & = & -\frac{\Gamma_{1\text{D}}}{\Gamma+2i\delta}e^{2ik_{0}(\text{z}_{0}-\text{z}')}\hat{a}_\text{in,R}(\text{z}'+v_g t),
\eea
where we have used that $\langle\hat{\sigma}_{00}(t)\rangle = \langle\hat{\sigma}_{00}(0)\rangle = 1$ for a emitter initially in the ground-state$|0\rangle$. We can do this because, once we eliminate the excited state the emitter can only be in the ground-state. For an emitter tuned into resonance $(\delta = 0)$ we get the well-known results of photon scattering in waveguides, with transmission and reflection amplitudes of $(1-\beta)$ and $\beta$, respectively \cite{Shen05}, where $\beta = \Gamma_{1\text{D}}/\Gamma$. This is illustrated in Fig. \ref{Spec1} (a) where we plot the transmitted intensity which shows a Lorentzian dip at resonance. The corresponding FWHM is found to be $\Gamma$. Thus, for a $1\text{D}$ waveguide with strong coupling to the emitter such that $\Gamma_{1\text{D}} \sim \Gamma$, scattering leads to complete reflection of the photon with the atom behaving as a mirror \cite{Shen07, Shen05, Zhou08}.
\subsection{A three-level emitter in V-configuration coupled to a one-dimensional waveguide}
Above we considered the simplest possible situation which could also easily be solved by other means. We now consider a situation, where the result is less obvious.  We choose an emitter in a V-configuration comprising a ground-state$\ket{0}$ and two excited states $\ket{1}$ and $\ket{2}$ located at some point $\text{z}_{0}$ in the waveguide (see Fig. \ref{fig2}~(b) for the schematic level structure). It is worth emphasizing that single photon scattering from such three-level emitters have been studied extensively in the past \cite{With10}. The purpose of addressing this problem here is to illustrate how the results of these previous works can be obtained directly with our method. To demonstrate the versatility of our approach, we assume that the exited states are coherently coupled by a (generally complex-valued) coupling $\Omega$. This then corresponds to a nonzero $\left[(\tilde{\mathcal{H}}_{\text{nh}})_{\text{nw}}\right] _{ee'}$ contribution to the non-Hermitian Hamiltonian $\tilde{\mathcal{H}}_{\text{nh}}$. Furthermore, we assume that the transitions from $\ket{0}$ to $\ket{1}$ and $\ket{0}$ to $\ket{2}$ are coupled to the waveguide mode with strengths $\mathcal{A}^{\mu =\zeta}_{1,(1\text{D})}$ and $\mathcal{A}^{\mu = \zeta}_{2,(1\text{D})}$ and decay with a total decay rate of $\Gamma_1$ and $\Gamma_2$ respectively. The Hamiltonian of the system is then given by $\hat{\mathcal{H}} = \hat{\mathcal{H}}_0 + \hat{\mathcal{V}} ~ (\hbar = 1)$ where,
\begin{align}
\label{eq26}
\hat{\mathcal{H}}_0 &= \sum_{j=0}^2 \omega_{jj}\hat{\sigma}_{jj}+ \Omega\hat{\sigma}_{12}+ \Omega^* \hat{\sigma}_{21}+ \hat{\mathcal{H}}_{F}
\\
\hat{\mathcal{V}} &= \sum_{j = 1}^{2}\sum_\mu\left(\mathcal{A}^{\mu}_{1}\hat{\sigma}_{0j}\hat{a}_{\mu} + \mathcal{A}^{\ast\mu}_{1}\hat{a}_{\mu}^\dagger~\hat{\sigma}_{j0}\right),
\end{align}
where as before we have defined $\hat{\sigma}_{ij} = |j\rangle \langle i|$.

The decay of the excited levels, $|1\rangle$ and $|2\rangle$ to modes other than the waveguide, is described by the Lindblad operators
\begin{align}
\label{eq27}
\hat{\mathcal{L}}_{s,1} &= \sqrt{\Gamma'_1} \hat{\sigma}_{10}
\\
\hat{\mathcal{L}}_{s,2} &= \sqrt{\Gamma'_2}\hat{\sigma}_{2 0}
\end{align}
with $\Gamma^{'}_{j}$ being the corresponding decay rate of the level $|j\rangle$. Note that as before, we have here used the relation $\Gamma^{'}_{j} = |\mathcal{A}^{\mu = s}_{j}|^{2}$ to define the decay rates out of the waveguide. The Lindblad operator for decay into the waveguide is given by 
\begin{align}
\label{eq28}
\hat{\mathcal{L}}_{\zeta_{(1\text{D})},1} &= e^{-i\phi_{1}}|\mathcal{A}^{\zeta}_{1,(1\text{D})}| ~\hat{\sigma}_{10},\nonumber
\\
& = e^{-i\phi_{1}}\sqrt{\Gamma^{\zeta}_{1, 1\text{D}}}~\hat{\sigma}_{10},
\\
\hat{\mathcal{L}}_{\zeta_{(1\text{D})}, 2} &= e^{-i\phi_{2}}|\mathcal{A}^{\zeta}_{2,(1\text{D})}| ~\hat{\sigma}_{20},\nonumber
\\
& = e^{-i\phi_{2}}\sqrt{\Gamma^{\zeta}_{2, 1\text{D}}}~\hat{\sigma}_{20}
\end{align}
In writing $\hat{\mathcal{L}}_{\zeta_{(1\text{D})},1}$ and $\hat{\mathcal{L}}_{\zeta_{(1\text{D})},2} $ in terms of the decay rates we have used the definition given in Eq. (\ref{eq14}), and introduced the phases $\phi_1$ and $\phi_2$ of the two couplings.

Now following Eq. (\ref{eq16}) - (\ref{eq18}), we set up the non-Hermitian Hamiltonian
\begin{align}
\label{eq29}
\hat{\mathcal{H}}_{\text{nh}} = \ \ &\tilde{\delta}_1 \hat{\sigma}_{11} + \tilde{\delta}_2  \hat{\sigma}_{22} + \mathcal{G}  \hat{\sigma}_{12} +\mathcal{G}^{\ast}  \hat{\sigma}_{21},
\end{align}
where we define the complex detunings $\tilde{\delta}_j = \delta_j - i \Gamma_j/2$ with $\delta_j = \omega_{jj} - \omega_{00} - \omega$ and $\Gamma_j = \Gamma'_j+\sum_{\zeta}\Gamma^{\zeta}_{j,1\text{D}}$, being the total line width of the excited state $|j\rangle$. Using Eq. (\ref{eq15}) and Eq. (\ref{eq18}) we can write a combined coupling term $\mathcal{G} = |\Omega|e^{i\theta}-i\sum_{\zeta}\sqrt{\Gamma^{\zeta}_{1,1\text{D}}\Gamma^{\zeta}_{2,1\text{D}}}e^{i(\phi_{1}-\phi_{2})}$. Note that due to the characteristic of the $\Gamma_{1\text{D}}$ coupling, the complex conjugation of the combined coupling gives $\tilde{\mathcal{G}} = |\Omega|e^{-i\theta}-i\sum_{\zeta}\sqrt{\Gamma^{\zeta}_{1,1\text{D}}\Gamma^{\zeta}_{2,1\text{D}}}e^{-i(\phi_{1}-\phi_{2})}$. Inversion of the non-Hermitian Hamiltonian in Eq. (\ref{eq29}) then yields
\begin{align}
\label{eq30}
\hat{\mathcal{H}}_{\text{nh}}^{-1} = \tilde{\delta}_{1,\rm eff}^{-1}  \hat{\sigma}_{11} + \tilde{\delta}_{2,\rm eff}^{-1}  \hat{\sigma}_{22} + \tilde{\mathcal{G}}_{\rm eff}^{-1}  \hat{\sigma}_{12} + \tilde{\mathcal{G}}_{\rm eff}^{'}  \hat{\sigma}_{21},
\end{align}
Here, we have written the inverse non-Hermitian Hamiltonian in terms of ``effective'' detunings and couplings
\begin{align}
\label{eq31}
\tilde{\delta}_{j,\rm eff} &= \tilde{\delta}_j - \frac{\mathcal{G}\tilde{\mathcal{G}}}{\tilde{\delta}_k}
\\
\label{eq32}
\tilde{\mathcal{G}}_{\rm eff} &= \frac{\mathcal{G}\tilde{\mathcal{G}} - \tilde{\delta_1} \tilde{\delta_2}}{\mathcal{G}},\\
\tilde{\mathcal{G}}^{'}_{\rm eff} &= \frac{\tilde{\mathcal{G}}\mathcal{G}- \tilde{\delta^\ast_1} \tilde{\delta^\ast_2}}{\tilde{\mathcal{G}}},
\end{align}
which depend both on the complex detunings of the excited states and on their couplings. The implications of these assignments will become more clear in the following. 

We first determine the output field using Eq. \eqref{eq5} at some spatial location $\text{z}$ to the right of the emitter,
\begin{align}
\label{eq33}
\hat{a}_{\text{out,R}}(\text{z}, t) & = \bigg[1+i\bigg\{\left(\Gamma^{\text{R}}_{1,1\text{D}}\right)\tilde{\delta}_{1,\rm eff}^{-1}+\left(\Gamma^{\text{R}}_{2,1\text{D}}\right)\tilde{\delta}_{2,\rm eff}^{-1}\nonumber
\\
&+\sqrt{\Gamma^{\text{R}}_{1, 1\text{D}}}(\tilde{\mathcal{G}}_{\rm eff}^{-1})\sqrt{\Gamma^{\text{R}}_{2, 1\text{D}}}e^{-i(\phi_1-\phi_2)}\nonumber\
\\
&+\sqrt{\Gamma^{\text{R}}_{2, 1\text{D}}}(\tilde{\mathcal{G}}_{\rm eff}^{'-1})\sqrt{\Gamma^{\text{R}}_{1, 1\text{D}}}e^{i(\phi_1-\phi_2)}\bigg\}\hat{\sigma}_{00}\bigg]\nonumber
\\
&\times\hat{a}_{\text{in,R}}(\text{z}-v_{\text{R}}t),
\end{align}
while the output field to the left of the emitter at some spatial location $\text{z}'$ is 
\begin{align}
\label{eq34}
\hat{a}_{\text{out,L}}(\text{z}', t) & = i\bigg[\sqrt{\Gamma^{\text{L}}_{1, 1\text{D}}}\tilde{\delta}_{1,\rm eff}^{-1}\sqrt{\Gamma^{\text{R}}_{1, 1\text{D}}}+\sqrt{\Gamma^{\text{L}}_{2, 1\text{D}}}\nonumber
\\
&\times\tilde{\delta}_{2,\rm eff}^{-1}\sqrt{\Gamma^{\text{R}}_{2, 1\text{D}}}+\sqrt{\Gamma^{\text{L}}_{1, 1\text{D}}}(\tilde{\mathcal{G}}_{\rm eff}^{-1})\sqrt{\Gamma^{\text{R}}_{2, 1\text{D}}}\nonumber\
\\
&\times e^{-i(\phi_1-\phi_2)}+\sqrt{\Gamma^{\text{L}}_{2, 1\text{D}}}(\tilde{\mathcal{G}}_{\rm eff}^{'-1})\sqrt{\Gamma^{\text{R}}_{1, 1\text{D}}}e^{i(\phi_1-\phi_2)}\bigg]\nonumber\
\\
&\times\hat{\sigma}_{00}e^{2ik_{0}(\text{z}_{0}-\text{z}')}\hat{a}_{\text{in,R}}(\text{z}'+v_{\text{L}}t).
\end{align}
Finding the photon scattering dynamics from even this relatively simple multi-level system is quite cumbersome, due to the complicated interplay of detunings and couplings. However, as can be seen from Eqs. (\ref{eq33}) and (\ref{eq34}), using the developed photon scattering formalism, we can straightaway provide a solution to even the general case in the limit of a single-photon/weak-field inputs. This is the key advantage of our formalism compared to many of the existing approaches \cite{Deutsch95, Chang12, Shi09, Shi11, Zheng11, Roy11, Laakso14, Fan10, Can15}.

From the above expressions we can see that the scattering amplitude strongly depends on the effective detunings $\tilde{\delta}_{\text{eff}}$ and the coupling $\tilde{\mathcal{G}}_{\text{eff}}$. Hence adjusting the quantities that appear in it, e.g., the coupling strength $\mathcal{G}$ between the excited states, it is possible to engineer this term to yield qualitatively different results. Thus one can invoke several different situations involving the emitter-waveguide coupling and the coupling between the excited states to analyze the behaviour of the output field further. To illustrate the dynamics, we restrict ourselves to the situation where the coupling is the same in both directions and the two-levels have the same decay rate. Thus, we consider $\Gamma^{\text{R/L}}_{1, 1\text{D}} = \Gamma^{\text{R/L}}_{2, 1\text{D}} = \Gamma_{1\text{D}}/2$ in Eqs. (\ref{eq33}) and (\ref{eq34}).  On eliminating the excited states the emitter can only be in the ground-state and hence for all later time $\langle\sigma_{00}\rangle = 1$. The output field at the right and left of the emitter is then given by
\bea
\label{35a}
\hat{a}_{\text{out,R}}(\text{z}, t) &=& \bigg[1+\frac{i\Gamma_{1\text{D}}}{2}\left(\frac{\tilde{\delta}+i\Gamma_{1\text{D}}-2|\Omega|\cos\Delta\phi}{\tilde{\delta}_{1}\tilde{\delta}_{2}-\mathcal{G}\tilde{\mathcal{G}}}\right)\bigg]\nonumber\\
&\times&\hat{a}_{\text{in,R}},\nonumber\\
\label{35b}
\hat{a}_{\text{out,L}}(\text{z}', t) &=& \frac{i|\Gamma_{1\text{D}}|}{2}\left(\frac{\tilde{\delta}+i\Gamma_{1\text{D}}-2|\Omega|\cos\Delta\phi}{\tilde{\delta}_{1}\tilde{\delta}_{2}-\mathcal{G}\tilde{\mathcal{G}}}\right)\nonumber\\
&\times&e^{2ik(\text{z}_{0}-\text{z}')}\hat{a}_{\text{in,R}}(\text{z}'+v_g t),
\eea
where $\tilde{\delta} = \tilde{\delta}_{1}+\tilde{\delta}_{2}$ and, $\Delta\phi = \theta-\left(\phi_{1}-\phi_{2}\right)$. We note here that the appearance of $\Delta\phi$ in these equations is a consequence of interferences between the different paths in Fig. \ref{fig2} (b). For instance level $|2\rangle$ can be reached by two different paths: either from direct excitation or through excitation to level $|1\rangle$ followed by transfer to level $|2\rangle$ by the coupling $\Omega$. These two paths interfere leading to the expressions above. 

From Eqs. (\ref{35a}) we see that by satisfying the condition $\tilde{\delta} =i\Gamma_{1\text{D}}-2|\Omega|\cos\Delta\phi$, the emitter can be made transparent to the incoming photon. This can be achieved by varying the phase and amplitude of the coherent coupling $\Omega$ which for example can be a magnetic field. We illustrate this in Fig. \ref{Spec1} (b), where we vary the drive phase $\phi$ and coupling $\Omega$ for fixed emitter parameters. Note that for the plot in Fig. \ref{Spec1} (b), we have assumed that the coupling strength of both the optical transitions are real. We also find that complete reflection from the emitter can occur under the condition $\delta_{1} = \delta_{2} = 0$, provided there is no loss to the outside of the waveguide and $\Gamma_{1\text{D}} \gg \Omega$. Thus we see that a three-level $V$ system can be made to selectively transmit or reflect a single photon thereby operating as a single-photon switch as required for transistors \cite{Kim98, Dayan08, Marco14, Sasha16}. 
\subsection{Scattering from multiple emitters coupled to a one-dimensional waveguide}
We next discuss the application of our photon scattering formalism to the case of multiple emitters coupled via the waveguide mode. We assume multi-level emitters to illustrate the full potential of our formalism. This problem is much more complicated in comparison to the ones we have discussed in the previous subsections. It however also contains rich physics due to quantum interference among various pathways of excitation and de-excitation. Additionally, it is also a prominent test bed for various interesting problems in quantum information sciences based on waveguide QED \cite{Peter_rmp}. As an example one can consider generation of entanglement between emitters over long distances via waveguide-mediated photons \cite{Gonzalez11}. Presently, established methods for solving such photon-scattering problem in multi-emitters system requires, setting up of a reduced master equation for the system and then performing numerical simulation to achieve the scattering amplitudes. In comparison, as will be shown in the following, one can find the scattering amplitudes directly using our photon scattering formalism.

\begin{figure}
	\centering
	\includegraphics[width = 0.5\textwidth]{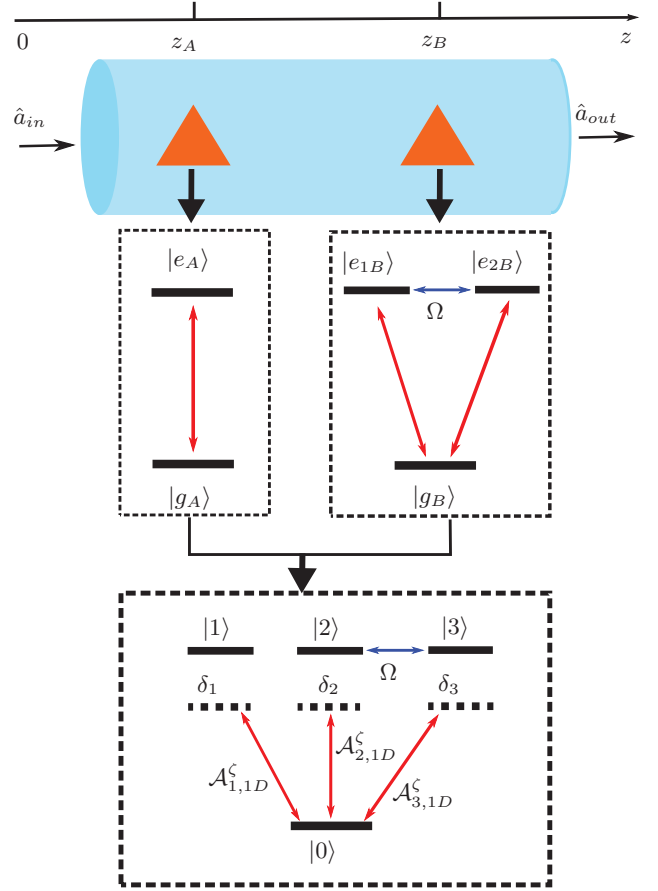}
	\caption{Two emitters in a waveguide (top) with individual level structures (center), and combined level structure in the single-excitation limit (bottom).\label{levelScheme1}}
\end{figure}

We begin our discussion with an example of two emitters coupled to a 1\text{D} mode of an optical waveguide. We label the two emitters as $\{A, B\}$ and consider them to be located at the spatial positions $\text{z}_{A}$ and $\text{z}_{B}$ respectively along the waveguide as shown schematically in Fig. \ref{levelScheme1}. The waveguide is assumed to be double-sided and we consider the input field (incident single-photon/weak coherent pulse) to be incident from the left and propagating to the right in the waveguide. 
We assume emitter $A$ to be a two-level system while emitter $B$ is a three-level V-type system, spaced $\Delta \text{z} = \text{z}_B-\text{z}_A$ apart. Emitter $A$ has ground-state$|g_A\rangle$ and excited state $|e_A\rangle$, whereas the three-level system $B$ consists of a single ground-state $|g_B\rangle$ and two excited states $|e_{1B}\rangle$ and $|e_{2B}\rangle$, coherently coupled at a rate $\Omega$ (for example with a magnetic field. For simplicity, we assume from now on that $\Omega = |\Omega|$ is real). The free Hamiltonian of this two-emitter system can be described as $(\hbar = 1)$
\begin{eqnarray}
\label{eq57}
\hat{\mathcal{H}}_{0}&=&\hat{\mathcal{H}}_{A_0}+\hat{\mathcal{H}}_{B_0}+\hat{\mathcal{H}}_{F}\\
\hat{\mathcal{H}}_{A_0}&=&\omega_{e,A}\hat{\sigma}^{A}_{ee}+\omega_{g,A}\hat{\sigma}^{A}_{gg}\\
\hat{\mathcal{H}}_{B_0}&=&\omega_{g,B}\hat{\sigma}^{B}_{gg}+\omega_{e1,B}\hat{\sigma}^{B}_{e_{1}e_{1}}+\omega_{e2,B}\hat{\sigma}^{B}_{e_{2}e_{2}}\nonumber\\
&+&\Omega\left(\hat{\sigma}^{B}_{e_{1}e_{2}}+\hat{\sigma}^{B}_{e_{2}e_{1}}\right),
\end{eqnarray}
where $\omega_i$'s are the free energies of the corresponding levels, $\hat{\mathcal{H}}_{F}$ is the standard free-field Hamiltonian and the atomic operators as before are defined by $\sigma_{ij} = |j\rangle\langle i|$. 

Our procedure is formulated in terms of the combined level structure of the emitters with one ground-state$|0\rangle\equiv|g_A, g_B\rangle$ and three excited states $|1\rangle\equiv|e_A, g_B\rangle$, $|2\rangle\equiv|g_A, e_{1B}\rangle$ and $|3\rangle\equiv|g_A, e_{2B}\rangle$ corresponding to a single excitation in either of the emitters as shown in Fig. \ref{levelScheme1}. In the combined basis we assume that the transitions from the ground levels to the excited levels $|1\rangle$, $|2\rangle$ and $|3\rangle$ are detuned from the incoming photon's frequency $\omega$ by $\delta_1$, $\delta_2$, and $\delta_3$ respectively. 

The interaction Hamiltonian $\hat{\mathcal{V}}$ describing the interaction of emitters with the photons in the combined basis $\{|0\rangle, |1\rangle, |2\rangle, |3\rangle\}$ is given by
\begin{eqnarray}
\label{eq57a}
\hat{\mathcal{V}}&=&\sum_{\mu}\mathcal{A}^{\mu}_{1,(1\text{D})} e^{i k_\mu \text{z}_A}(\hat{a}_{\mu}^{\dagger}|0\rangle\langle 1|+|1\rangle\langle 0|\hat{a}_{\mu})\nonumber\\
&+&\sum^{3}_{j = 2}\sum_{\mu}e^{i k_\mu \text{z}_B}\mathcal{A}^{\mu}_{j,(1\text{D})}(\hat{\sigma}_{0j}\hat{a}_{\mu}+\hat{a}_{\mu}^{\dagger}\hat{\sigma}_{j0})
\end{eqnarray}
where the dipole transitions between the states $|j\rangle (j = 1,2,3)$ and $|0\rangle$ are coupled to the waveguide mode with strengths $\mathcal{A}^{\zeta}_{j, (1\text{D})}$ respectively. We assume these couplings have no additional phase (such that $\mathcal{A}^{\zeta}_{j, (1\text{D})}$ is real-valued) apart from the phase contribution originating from the distinct positions of the emitters in the waveguide, $e^{i k_0 z_{A/B}}$. As a result of these phases the incoming field couples to emitter $B$ with an additional phase $e^{i k_0 (\text{z}_B-\text{z}_A)}$ relative to the field at position $\text{z}_A$. Ignoring an overall phase, we from this point assume emitter $A$ as the reference point $\text{z}_A=0$ and as such $\text{z}_B=\Delta\text{z}$. Note that, in writing Eq. (\ref{eq57a}) we have assumed that the spatial separation of the emitters $\Delta \text{z}$ is much larger than the wavelength $\lambda$ of the incoming photon. We have therefore ignored the possibility of any direct interaction (like dipole-dipole) between the emitters and focus only on the waveguide-mediated interaction. We do, however, explicitly include such direct interaction and discuss their influence on the emitter dynamics towards the end of this section.

The Hamiltonian of the combined system can then be written as $\hat{\mathcal{H}}=\hat{\mathcal{H}}_0+\hat{\mathcal{V}}$, where now
\begin{equation}
\label{eq58}
\hat{\mathcal{H}}_0=\sum_{i=0}^{3}\omega_{ii}|i\rangle\langle i|+\Omega\left(\hat{\sigma}_{23}+\hat{\sigma}_{32}\right)+\hat{\mathcal{H}}_{F}.
\end{equation}
Based on this full Hamiltonian $\mathcal{H}$, we next wish to construct the excited-subspace Hamiltonian $\tilde{\mathcal{H}}_{\text{nh}}$ similar to Eq. (\ref{eq18}) in the basis $(|1\rangle,|2\rangle,|3\rangle)$. For this purpose we need to consider the decays of the excited state, which in this case is represented by the Lindblad operators 
\bea
\label{eq58a}
\hat{\mathcal{L}}_{s_j} & = &\sqrt{\Gamma^{'}_{j}}~\hat{\sigma}_{j0}\\
\hat{\mathcal{L}}_{\zeta_{(1\text{D})},j}  & = &\sqrt{\Gamma^{\zeta}_{j,1\text{D}}}~\hat{\sigma}_{j0}, 
\eea
where as before $\Gamma^{'}_{j}$ is the decay rate of state $|j\rangle$ out of the waveguide, while $\Gamma^{\zeta}_{j,1\text{D}}$ is the decay rate into the waveguide along the direction $\zeta$. Note that, in writing the expression of $\hat{\mathcal{L}}_{\zeta_{(1\text{D})}, j}$ we have used the definition in Eq. (\ref{eq14}). 

Taking into consideration all of these terms the diagonal part of the non-Hermitian Hamiltonian becomes
\bea
(\hat{\tilde{\mathcal{H}}}_{\text{nh}})_{d} = \sum_{j=1}^3 \tilde{\delta}_j \hat{\sigma}_{jj}
\eea
where the complex detuning $\tilde{\delta}_j=\delta_j-\frac{i\Gamma_j}{2}$, with $\Gamma_j = \Gamma_{j,1\text{D}}+\Gamma^{'}_j$ being the total decay rate of transition $|j\rangle\rightarrow|0\rangle$. Here the decay into the waveguide is defined as before $\Gamma_{j,1\text{D}} = \sum_\zeta \Gamma^{\zeta}_{j,1\text{D}}$. The detuning is defined as $\delta_j= (\omega_{jj}-\omega_{00}-\omega)$, where $\omega$ is the central frequency of the incoming photon.

We next construct the off-diagonal part of the non-Hermitian Hamiltonian of the combined system $\hat{\tilde{\mathcal{H}}}_{\text{nh}}$. To simplify this Hamiltonian we make an assumption about the nature of coupling between the emitters and the waveguide mode. We assume that the coupling strengths are the same along both the propagation directions, i.e., $\mathcal{A}^{(R)}_{j,(1\text{D})} = \mathcal{A}^{(L)}_{j,(1\text{D})} = \mathcal{A}_{j,(1\text{D})}$. Using Eq. (\ref{eq15}) and Eq. (\ref{eq18}) we then find that the off-diagonal elements of $\hat{\tilde{\mathcal{H}}}_{\text{nh}}$ consist of the waveguide-mediated interaction terms of the form,
\begin{eqnarray}
\label{eq59}
(\hat{\tilde{\mathcal{H}}}_{\text{nh}})_{\text{w}}&=&-\frac{i}{2}\sqrt{\Gamma_{1,1\text{D}}\Gamma_{2, 1\text{D}}}e^{i k \Delta \text{z}}(\hat{\sigma}_{21}+\hat{\sigma}_{12})\nonumber\\
&&-\frac{i}{2}\sqrt{\Gamma_{1,1\text{D}}\Gamma_{3,1\text{D}}}e^{i k \Delta \text{z}}(\hat{\sigma}_{31}+\hat{\sigma}_{13})\nonumber\\
&&-\frac{i}{2}\sqrt{\Gamma_{2,1\text{D}}\Gamma_{3,1\text{D}}}(\hat{\sigma}_{32}+\hat{\sigma}_{23})
\end{eqnarray}
and the non-waveguide couplings, which in this case is just the coherent coupling $\Omega$
\begin{equation}
(\hat{\tilde{\mathcal{H}}}_{\text{nh}})_{\text{nw}} = \frac{\Omega}{2}\left(\hat{\sigma}_{23}+\hat{\sigma}_{32}\right).
\end{equation}
Note that in writing Eq. (\ref{eq59}) we have used the definition of $\Gamma_{j, 1\text{D}}$ in terms of the coupling strengths from Eq. (\ref{eq14}).
\begin{figure}
\includegraphics[height = 6.5 cm]{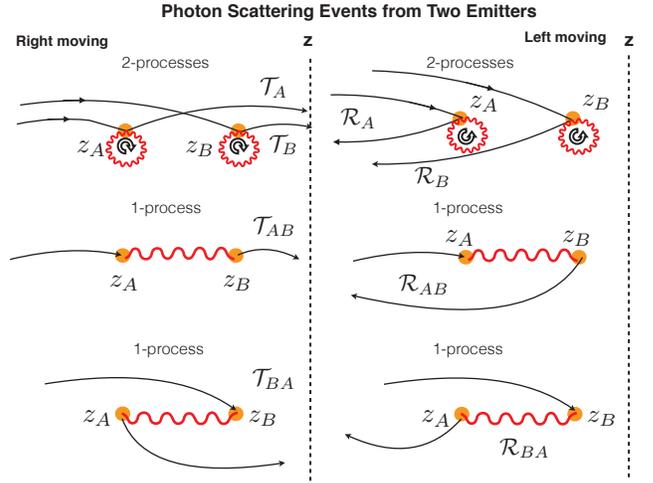}
\caption{Schematic of light scattering from two generic emitters located at the position $\text{z}_{A}$ and $\text{z}_{B}$ in a double-sided waveguide with a right-going input photon pulse . Here $\mathcal{T}_{i}$ and $\mathcal{R}_{i}$ signifies the single emitter transmitted and reflected amplitudes respectively. Amplitude for transmitted and reflected light for scattering involving two emitters are on the other hand given by $\mathcal{T}_{ij}$ and $\mathcal{R}_{ij}$, respectively. The wiggly lines signify field-mediated interactions between the emitters in terms of the non-Hermitian Hamiltonian $\tilde{\mathcal{H}}_{\text{nh}}$ as discussed in the text. The wiggly circles with arrows inside symbolizes the scattering event.}
\label{figure3}
\end{figure}
Finally, we arrive at the non-Hermitian Hamiltonian
\begin{equation}
\label{eq61}
\tilde{\mathcal{H}}_{\text{nh}} = \left(\begin{array}{ccc} \tilde{\delta}_1  & -\frac{i}{2}\Gamma_{12} & -\frac{i}{2}\Gamma_{13} \\
-\frac{i}{2}\Gamma_{12}&\tilde{\delta}_2 &(\frac{\Omega}{2}-\frac{i}{2}\Gamma_{23}) \\
-\frac{i}{2}\Gamma_{13}&(\frac{\Omega}{2}-\frac{i}{2}\Gamma_{23}) & \tilde{\delta}_3\\
\end{array}\right),
\end{equation}
in the excited subspace defined by the basis $(|1\rangle,|2\rangle,|3\rangle)$. Here we have defined complex couplings $\Gamma_{12}=\sqrt{\Gamma_{1,1\text{D}}\Gamma_{2,1\text{D}}}e^{i k \Delta \text{z}}$, $\Gamma_{13}=\sqrt{\Gamma_{1,1\text{D}}\Gamma_{3, 1\text{D}}}e^{i k \Delta \text{z}}$ and $\Gamma_{23}=\sqrt{\Gamma_{2, 1\text{D}}\Gamma_{3,1\text{D}}}$. Next, on taking inverse of Eq. (\ref{eq61}) we get 
\begin{equation}
\label{eq62}
[\tilde{\mathcal{H}}_{\text{nh}}]^{-1}=\left(\begin{array}{ccc}
{\delta}_{1,\text{eff}}^{-1}  & \Gamma_{12,\text{eff}}^{-1} & \Gamma_{13,\text{eff}}^{-1} \\
\Gamma_{12,\text{eff}}^{-1}&{\delta}_{2,\text{eff}}^{-1} &\Gamma_{23,\text{eff}}^{-1} \\
\Gamma_{13,\text{eff}}^{-1}&\Gamma_{23,\text{eff}}^{-1} & {\delta}_{3,\text{eff}}^{-1}\\
\end{array}\right)
\end{equation}
where the effective detunings and couplings are defined in Appendix C. 

We next study the scattering of a single-photon pulse. In Fig. \ref{figure3} we sketch the different possible scattering processes involved for a two-emitter system. As can be seen from  Fig. \ref{figure3} there are several processes to account for. Our formalism, however, is well equipped to handle such complications and the photon-scattering relation stated in Eq. (\ref{eq5}) can straightaway give the solution to this scattering problem. Conveniently the multiple scattering pathways can be simply written as a matrix multiplication between the vectors $\mathcal{V}_\pm$ and the matrix $\mathcal{\tilde{H}}_{\text{nh}}^{-1}$. If we come with a right-going input field from the left, the total outgoing field to the right of the emitters is then following Eq. (\ref{eq5}), given by
\bea
\label{eq65}
\hat{a}_{out,\text{R}}(\text{z},t) & = &\Bigg[1+i\Bigg(\T_A+\T_B+\T_{AB,12}+\T_{BA,12}\nonumber\\
&+&\T_{AB,13}+\T_{BA,13}\Bigg)\hat{\sigma}_{00}\Bigg]\hat{a}_{in,R}(\text{z}-v_g t),\nonumber\\
\eea
where we have divided all possible scattering pathways into separate parts with their respective transition amplitudes $\T$. These are expressed using the elements of the non-Hermitian Hamiltonian in Eq. (\ref{eq62}), and are given by 
\bea
\label{eq63}
&&\T_A=\frac{\Gamma_{1,1\text{D}}}{2\delta_{1,\text{eff}}},\nonumber\\
&&\T_B=\frac{\Gamma_{2,1\text{D}}}{2\delta_{2,\text{eff}}}+\frac{\Gamma_{3, 1\text{D}}}{2\delta_{3,\text{eff}}}+\frac{\sqrt{\Gamma_{2, 1\text{D}}\Gamma_{3, 1\text{D}}}}{\Gamma_{23,\text{eff}}},\nonumber\\
&&\T_{AB,12}+\T_{BA,12}=\frac{\sqrt{\Gamma_{1,1\text{D}}\Gamma_{2, 1\text{D}}}}{\Gamma_{12,\text{eff}}}\cos(k_0 \Delta\text{z}),\nonumber\\
&&\T_{AB,13}+\T_{BA,13}=\frac{\sqrt{\Gamma_{1, 1\text{D}}\Gamma_{3, 1\text{D}}}}{\Gamma_{13,\text{eff}}}\cos(k_0 \Delta\text{z}).
\eea 
Note that in writing Eq. (\ref{eq65}) we have neglected the noise as the photon at output is typically detected in photodetectors where the noise owing to vacuum does not contribute. From Eqs. (\ref{eq65}) and (\ref{eq63}), we find that owing to the scattering from the two emitters the amplitudes now contain some interference terms $\cos(k_0 \Delta \text{z})$ depending on the emitter separation. 

To investigate the characteristic of the outgoing field further, we below consider some specific cases with respect to the emitter configurations and couplings. We assume that initially both the emitters are in their ground-states. Similar to above we can then replace the ground-state operator $\hat{\sigma}_{00}$ by $\langle\hat{\sigma}_{00}\rangle = 1$, since the combined system only has a single ground-state after elimination of the excited states. 

\subsubsection{Two Two-Level emitters} 
As a first example let us consider emitter $B$ to behave effectively as a two-level system. This can happen if the transition $|3\rangle \rightarrow |0\rangle$ does not couple to the waveguide mode such that $\Gamma_{3, 1\text{D}}=0$ and $|3\rangle$ also does not couple coherently to any other level of emitter $B$, i.e., $\Omega=0$. Then, the total right-going output field for a single right-going input field coming from the left is reduced to
\begin{multline}
\label{eq67}
\hat{a}_{out,\text{R}}(\text{z},t)=\Big[1+i\Big(\frac{\Gamma_{1,1\text{D}}}{2\delta_{1,\text{eff}}}+\frac{\Gamma_{2, 1\text{D}}}{2\delta_{2,\text{eff}}}\\+\frac{\sqrt{\Gamma_{1, 1\text{D}}\Gamma_{2, 1\text{D}}}}{\Gamma_{12,\text{eff}}}\cos(k \Delta \text{z})\Big)\Big]\hat{a}_{in,\text{R}}(\text{z}-v_g t),
\end{multline}
while the reflected field is given by 
\begin{multline}
\label{eq68}
\hat{a}_{out,\text{L}}(\text{z}',t)=i\Big[\frac{\Gamma_{1, 1\text{D}}}{2\delta_{1,\text{eff}}}e^{2 i k (\text{z}_A-\text{z})}+\frac{\Gamma_{2, 1\text{D}}}{2\delta_{2,\text{eff}}}e^{2 i k (\text{z}_B-\text{z})}\\+\frac{\sqrt{\Gamma_{1, 1\text{D}}\Gamma_{ 2, 1\text{D}}}}{\Gamma_{12,\text{eff}}}\cos(k \Delta \text{z})\Big]\hat{a}_{in,\text{R}}(\text{z}'+v_g t).
\end{multline}
If we next assume that the emitters are identical, i.e., $\Gamma_{i, 1\text{D}} = \Gamma_{1\text{D}}$, $\Gamma_{1}=\Gamma_{2}\equiv\Gamma_{1\text{D}}+\Gamma'$ and $\delta_1=\delta_2\equiv \delta$, we can after some simplifications find the transmitted output field to be,
\bea
\label{eq69}
\hat{a}_{out,\text{R}} & = & \bigg[1-\frac{2\Gamma_{1\text{D}}+(1-e^{2i k \Delta \text{z}})\frac{\Gamma_{1\text{D}}^2}{(\Gamma'+2i\delta)}}{(\Gamma'+2\Gamma_{1\text{D}}+2i\delta)+(1-e^{2i k \Delta \text{z}})\frac{\Gamma_{1\text{D}}^2}{(\Gamma'+2i\delta)}}\bigg]\nonumber\\
&&\times\hat{a}_{in,\text{R}}(\text{z}-v_{g}t).
\eea
The transmission spectrum evaluated from Eq. (\ref{eq69}) can be shown to be similar to that of a cavity of length $L = \Delta \text{z} = (\text{z}_B-\text{z}_A)$. Furthermore, for $\Delta \text{z} = q\lambda/2$, where $\lambda$ is the wavelength of the incoming photon and $q$ is an integer, the transmitted amplitude is given by 
\begin{equation}
\label{eq70}
\hat{a}_{out,\text{R}} = \left[1-\frac{2\Gamma_{1\text{D}}}{\Gamma'+2\Gamma_{1\text{D}}+2i\delta}\right]\hat{a}_{in,\text{R}}
\end{equation}
From the above expression it is clearly visible that the system of two emitters become perfectly reflective at resonance and for $\Gamma' = 0$. The transmission spectrum then has a  Lorentzian window with a width twice that of a single two-level system, due to the effective enhancement of $\Gamma_{1\text{D}}$ as compared to Eq. (\ref{eq24}) for a single two-level emitter. We find that the emitter system thus behaves as an `atomic mirror' with $N_A=2$. This problem was also investigated in Ref. \cite{Chang12} where the phenomenon of an atomic mirror was reported for multiple emitters. We immediately obtain the same result as \cite{Chang12} by our formalism, thus exhibiting the strength and simplicity of it. 

Additionally, one finds that for emitter spacings close to $\sin(k\Delta \text{z})\approx0$, the spectrum contains an ultra-narrow transparency window at $\delta\approx\frac{\Gamma_{1\text{D}}}{2}\sin(k\Delta \text{z})$. Thus, the system moves away from behaving like a mirror with minor change in $\Delta \text{z}$ about $\Delta\text{z} = n\lambda/2$. This can be understood from the fact that the dark state, which was in resonance with the bright state, gets shifted by $\delta$ and starts to couple to light. We find that the FWHM of the resonance line due to the dark state is now given by $\Gamma_{1\text{D}}\sin^{2}(k\Delta z)/2$. Note that in principle this could be used to transform the waveguide-emitter system into a narrow frequency filter that selectively allow photons to pass through for suitable separation distance between the emitters. The change in the separation can be introduced via external control, for example by moving atoms trapped near a waveguide.

Alternatively, for $\Delta \text{z} = (2q+1)\lambda/4$, the transmitted amplitude becomes
\bea
\label{eq70a}
\hat{a}_{out,\text{R}} & = &\frac{(\Gamma'+2i\delta)^{2}}{(\Gamma'+2i\delta)^{2}+2\Gamma_{1\text{D}}(\Gamma'+2i\delta+\Gamma_{1\text{D}})}\nonumber\\
&\times& \hat{a}_{in,\text{R}}(\text{z}-v_{g}t).
\eea
In this case one finds that the transmission spectrum for $\Gamma' = 0$ has a window at resonance with a width $\sqrt{2}\Gamma_{1\text{D}}$. 
\begin{figure}
\begin{tabular}{ccc}
	\includegraphics[width = 0.22\textwidth]{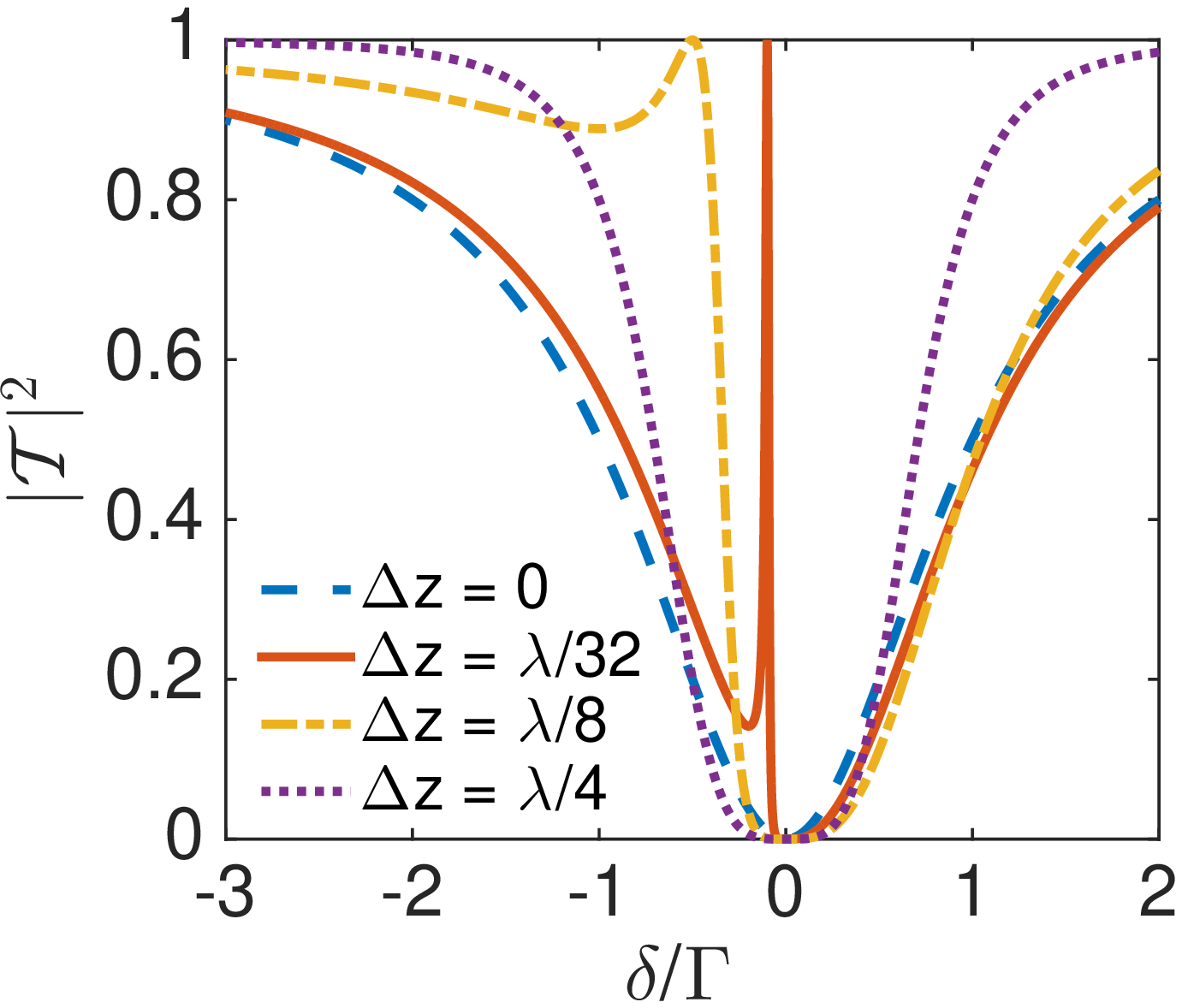} & \includegraphics[width = 0.25\textwidth]{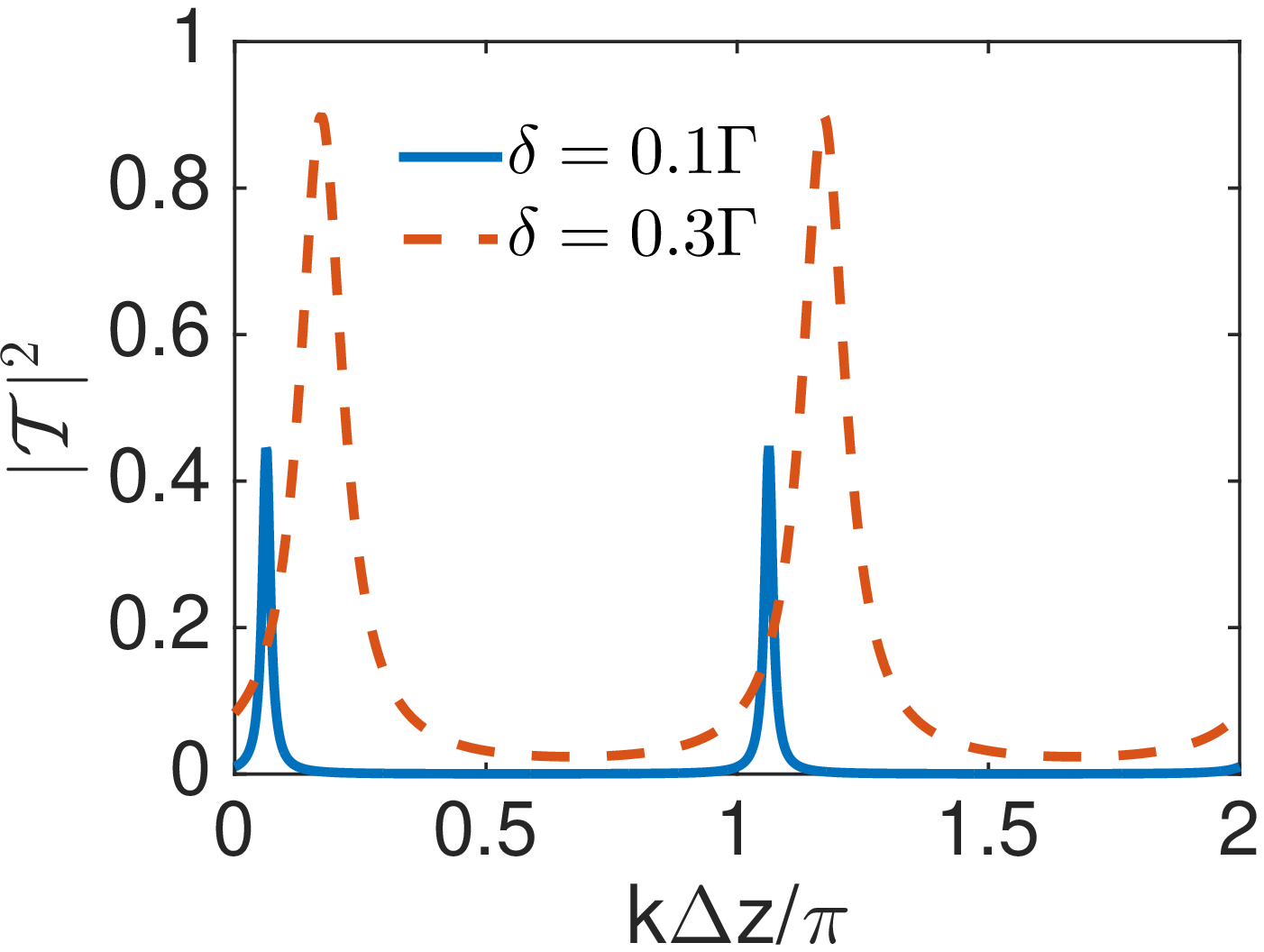}\\
	(a) & (b) 
\end{tabular}	
	\caption{Transmission $|\mathcal{T}|^2=|\langle \hat{a}^\dagger_{out,R} \hat{a}_{out,R}\rangle/\langle \hat{a}^\dagger_{in} \hat{a}_{in}\rangle|$ for two two-level emitters coupled to a 1\text{D} waveguide. The parameters used for the plots are (a) $\beta=1$ and comparing four different values of the phase distance $k\Delta z$, (b)  transmission as a function of the phase distance $k\Delta z$ for $\delta_1=\delta_2=\delta = 0.1\Gamma$ and $0.3\Gamma$, $\beta=0.99$.. }
	\label{Spec3}		
\end{figure}

In Fig. \ref{Spec3} (a) using Eq. (\ref{eq69}) we show the transmitted intensity for the two-emitter system as a function of the detuning. In Fig. \ref{Spec3} (b) we show the transmitted intensity for the two-emitter system for varying spacings of the emitters. The transmission resonances arise from the fact that the dark state starts to resonantly couple to the light field. 

In the above discussion, we have only considered interactions between the emitters mediated by the waveguide. In the following, we address the question of closely spaced emitters interacting with each other via their dipolar fields. For $\Delta \text{z} \leq \lambda$, there is strong dipole-dipole interaction between the emitters \cite{Dicke, Scully06, Das08, Das10, Cheng17} and the off-diagonal term in the non-Hermitian Hamiltonian of Eq. (\ref{eq62}) is thus modified. In addition to the waveguide-mediated coupling, these terms will have contributions from the direct dipole-dipole interactions $\mathcal{V}_{AB} (\mathcal{V}_{BA})$ between the optical transitions of the emitters along with collective decays $\Gamma'_{c}$ to the outside. In the limit of very small separation, where we can neglect the phase difference from propagation, the two-emitter system in the single-excitation regime effectively reduces to a single three-level system with dynamics similar to that discussed before in Sec. IV.A. Here the effective V-configuration is realized by defining a symmetric and anti-symmetric state which are the eigen-basis of the dipole-coupling Hamiltonian. Here we shall consider how this situation emerges from the single excitation subspace spanned by the basis $\{|e_{A},g_{B}\rangle, |g_{A},e_{B}\rangle \}$ of the emitters $A$ and $B$. As such, the subscripts $1$ and $2$ in Eq. (\ref{eq61}) in the previous case are now replaced with $A$ and $B$ respectively. The non-Hermitian Hamiltonian then becomes
\begin{equation}
\mathcal{H}_{\text{nh}}=
\left[\begin{array}{cc} \tilde{\delta}_A  & \mathcal{V}_{AB}-\frac{i}{2}\sqrt{\Gamma_{A, 1\text{D}}\Gamma_{B, 1\text{D}}}\\
\mathcal{V}_{BA}-\frac{i}{2}\sqrt{\Gamma_{A, 1\text{D}}\Gamma_{B,1\text{D}}}  & \tilde{\delta}_B
\end{array}\right]
\end{equation}
where compared to Eq. (\ref{eq61}) we now have an extra off-diagonal elements describing the direct dipole-dipole interaction between the two closely separated emitters. As before we define the complex detuning $\tilde{\delta}_j =(\delta_j-\frac{i}{2}\Gamma_{j})$ with $( j = A, B)$ and the total decay rate of each emitter given by $\Gamma_{j}  = \Gamma^{'}_{j}+\Gamma_{j, 1\text{D}}$. Furthermore, in this case we consider the limit $k\Delta z\rightarrow 0$ for waveguide-mediated coupling. Using this we find the transmitted field to be
\bea
\label{eq71}
&&\hat{a}_{out,\text{R}} = \bigg\{1+\bigg[4i\sqrt{\Gamma_{A, 1\text{D}}\Gamma_{B, 1\text{D}}}|\mathcal{V}|\cos\phi+2\Gamma_{A,1\text{D}}\nonumber\\
&&\Gamma_{B, 1\text{D}}-\Gamma_{A, 1\text{D}}(\Gamma_B+2i\delta_B)-\Gamma_{B, 1\text{D}}(\Gamma_A\nonumber\\
&&+2i\delta_A)]\bigg/\bigg[(\Gamma_A+2i\delta_A)(\Gamma_B+2i\delta_B)-\Gamma_{A,1\text{D}}\Gamma_{B, 1\text{D}}\nonumber\\
&&-4i\sqrt{\Gamma_{A, 1\text{D}}\Gamma_{B, 1\text{D}}}|\mathcal{V}|\cos\phi+4|\mathcal{V}|^2)\bigg]\bigg\}\hat{a}_{in}(\text{z}-v_{g}t).\nonumber\\
\eea
Here we have assumed that the dipole interaction between the emitters has the form $\mathcal{V}_{AB} = (\mathcal{V}_{BA})^\ast = |\mathcal{V}|e^{i\phi}$.

\subsubsection{A two-level and a three-level emitter} 
Let us now investigate how the coherent coupling between level $|2\rangle$ and $|3\rangle$ of the second emitter influences the scattering dynamics. The effect of interference due to such coherent coupling is different than that due to the waveguide mediated coupling. To elaborate further, let us compare the two two-level emitter case with the present situation where the coherent coupling is non-zero, $\Omega\neq0$. Following two two-level emitter example we now assume $\delta_2=\delta_3=\delta_B$, $\delta_1=\delta_A$, $\Gamma_{1, 1\text{D}}=\Gamma_{2, 1\text{D}} = \Gamma_{3, 1\text{D}} = \Gamma_{1\text{D}}$ and all $\Gamma'_i=0$. The transmitted field is then given by 
\bea
\label{eq72}
\hat{a}_{out,\text{R}}(\text{z},t) & = & \frac{2\delta_A(\Omega+\delta_B)}{e^{2i k \Delta z}\Gamma_{1\text{D}}^2-(\Gamma_{1\text{D}}+2i\delta_A)[\Gamma_{1\text{D}}+i(\Omega+\delta_B)]}\nonumber\\
&\times&\hat{a}_{in, \text{R}}(\text{z}-v_g t).
\eea

\begin{figure}
\begin{tabular}{ccc}
	\includegraphics[width = 0.23\textwidth]{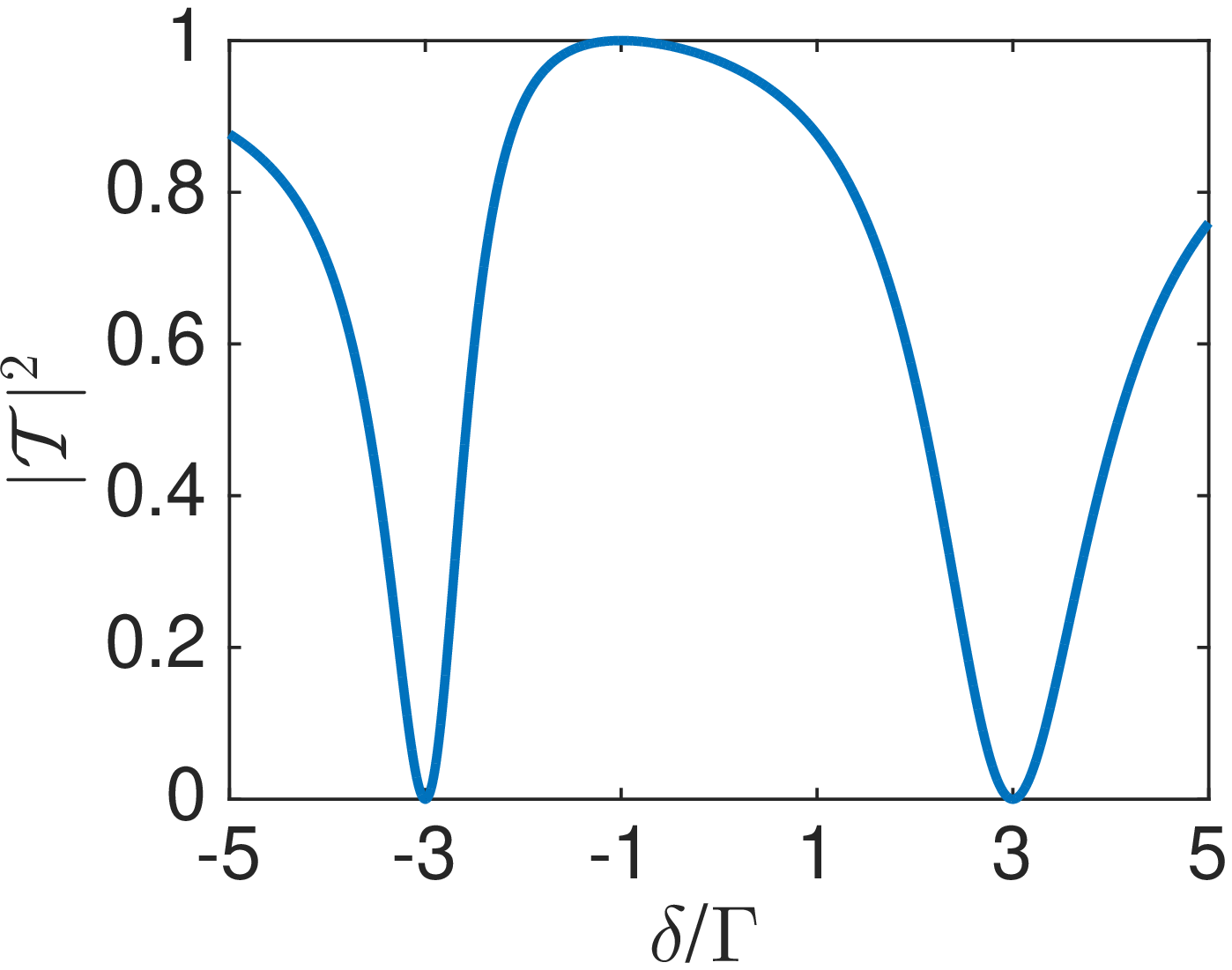} & \includegraphics[width = 0.25\textwidth]{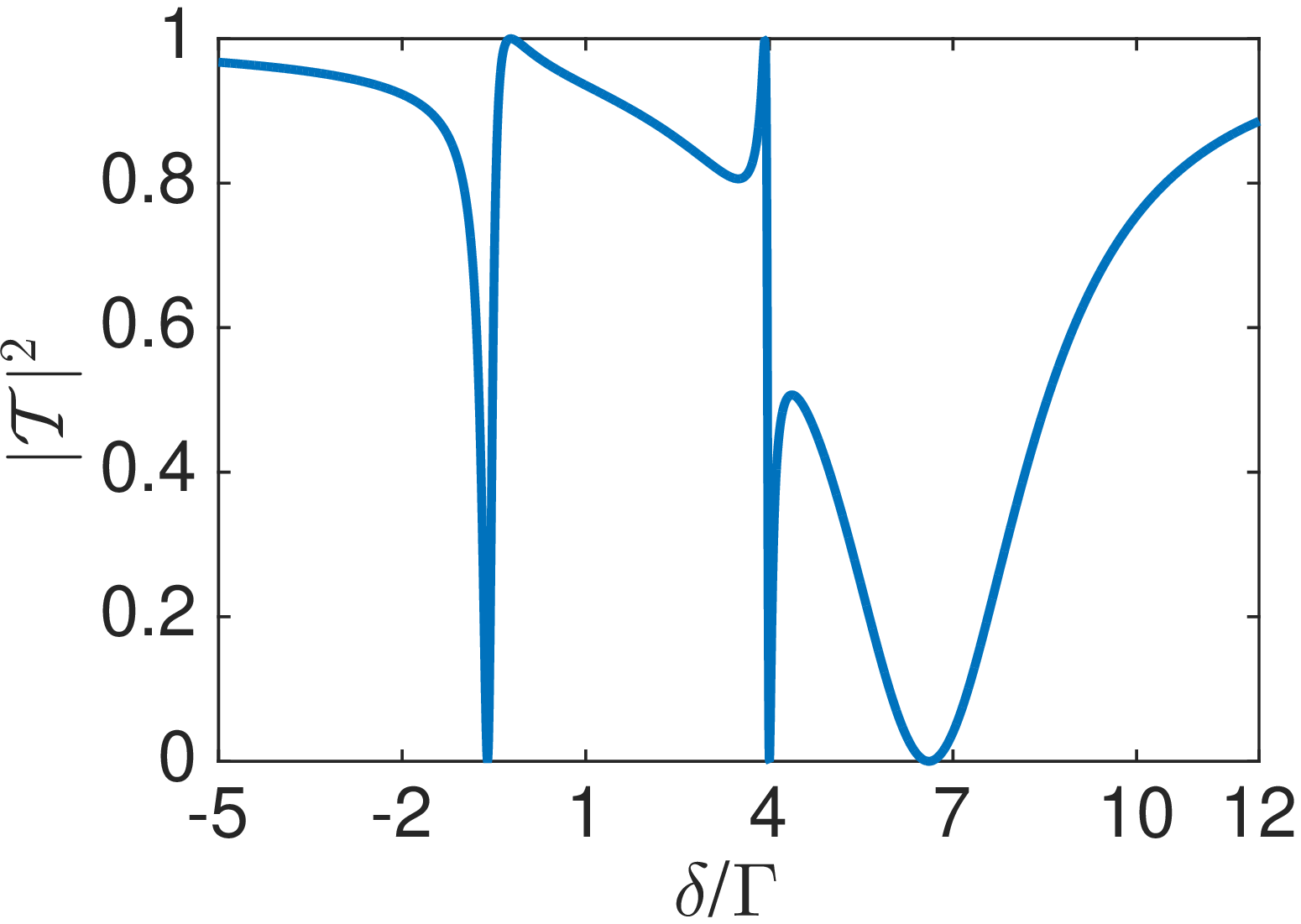}\\
	(a) & (b)
\end{tabular}
\caption{Transmission $|\mathcal{T}|^2=|\langle \hat{a}^\dagger_{out,R} \hat{a}_{out,R}\rangle/\langle \hat{a}^\dagger_{in} \hat{a}_{in}\rangle|$ from a two-emitter system. Here we consider a combination of a two-level emitter and a three-level emitter in the V-configuration coupled to a 1\text{D} waveguide. The parameters used for the plots are as follows, for  (a) $\delta_A=-3\Gamma-\delta$. $\delta_B=-2\Gamma-\delta$ and $\beta = 1$, $k\Delta z=2\pi$ and $\Omega=5$ while for (b) $\delta_1=4\Gamma-\delta$, $\delta_2= -\delta$, $\delta_3=6\Gamma-\delta$, $\Gamma_{1, 1\text{D}}= 0.1\Gamma$, $\Gamma_{2, 1\text{D}}= \Gamma$, $\Gamma_{3, 1\text{D}} = 3\Gamma$, $k\Delta z=1$ and $\Omega=2$.}
\label{Spec5}
\end{figure}
We show the transmission spectrum evaluated using Eq. (\ref{eq72}) in Fig. \ref{Spec5} (a) . We find that the transmission spectrum has two points of total reflection: at resonance with emitter $A$, i.e., $\delta_A = 0$ and at $\delta_B=-\Omega$. At $\delta_A = 0$, the input photon is completely reflected off the emitter $A$ which behaves as a perfect mirror and thus emitter $B$ does not `see' any input field. The scattered output field from the two-emitter system thus has characteristics reminiscent of total reflection off a single two-level emitter. The width of this resonance is $\Gamma$. At $\delta_B=-\Omega$, the incoming field is in resonance with the symmetric state, an eigenstate of emitter $B$'s excited-subspace Hamiltonian. From Eq. (\ref{eq72}) we find the width of this resonance to be $2\Gamma$. 

Finally, our method allows evaluating the scattering dynamics for a general emitter system. We give an example of this in Fig. \ref{Spec5} (b) which displays a complex interplay between various processes.  
\section{Application of the photon scattering formalism to emitters with two or more ground-states}
Until now we have discussed examples that involve only a single ground-state. Thus, we have not yet needed the effective operator master equation. To illustrate the full use of our formalism, in this section we solve a scattering problem involving an emitter with multiple ground-states. We will first introduce the model system in Sec. V.A and discuss the relevant Hamiltonian and equation of motions. Then in Sec. V.B and in the subsequent subsections, we discuss in detail the scattering dynamics of a single photon and a weak coherent pulse. 
\begin{figure}
\centering
\includegraphics[width=0.3\textwidth]{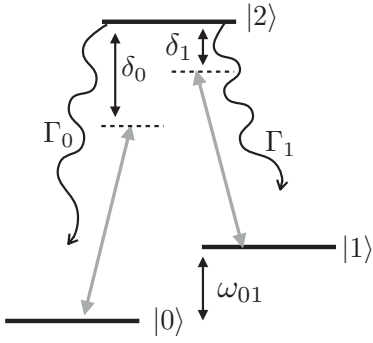}
\caption{Three-level emitter with a $\Lambda$-type level structure consisting of two groundstates ($|0\rangle$, $|1\rangle$) and one excited state ($|2\rangle$).}
\label{lambdasys}
\end{figure}
\subsection{The Model System and Hamiltonian}
For this purpose we consider a single three-level $\Lambda$-type emitter coherently coupled to a waveguide as shown schematically in Fig. \ref{lambdasys}. Such a system is generally described by a Hamiltonian $\hat{\mathcal{H}} = \hat{\mathcal{H}}_0+\hat{\mathcal{V}} (\hbar = 1)$ where,
\begin{subequations}
\begin{equation}
\hat{\mathcal{H}}_0=\sum_{j=0}^{2} \omega_{jj} \hat{\sigma}_{jj}+\hat{\mathcal{H}}_{F}
\end{equation}
\begin{equation}\label{Vpequation}
\hat{\mathcal{V}} = \sum_\mu\mathcal{A}^{\mu}_{0} \hat{a}_{\mu}^{\dagger} \hat{\sigma}_{20}+\sum_\mu\mathcal{A}^{\mu}_{1} \hat{a}_{\mu}^{\dagger} \hat{\sigma}_{21} + H.c.
\end{equation}
\end{subequations}
Here $\hat{\mathcal{H}}_0$ and $\hat{\mathcal{H}}_{F}$ are the free-energy and free-field Hamiltonian, respectively, while the excitation (de-excitation) operators are defined by $\hat{\mathcal{V}}_{+} (\hat{\mathcal{V}}_{-} = [\hat{\mathcal{V}}_{+}]^\dagger)$. The frequencies $\omega_{jj}$ correspond to the energies of levels $|j\rangle$. We assume that the emitter transitions $|2\rangle\rightarrow|j\rangle$ couples to the $1$D waveguide mode with a coupling strengths $\mathcal{A}^{\mu = \zeta}_{j,(1\text{D})}$ and $\hat{a}_{\mu = \zeta}$ ($\hat{a}_{\mu = \zeta}^{\dagger}$) represent the corresponding annihilation (creation) operator of the waveguide mode. Considering the coupling strengths to be real we can then, following Eq. (\ref{eq14}), write the decay from $|j\rangle$ into the waveguide as $ \Gamma_{j,1\text{D}}^{\zeta} =(\mathcal{A}^{\zeta}_{j,(1\text{D})})^2$. Furthermore, the decay to the outside of the waveguide is as before, given by $\Gamma^{'}_{j} = (\mathcal{A}^{\mu = s}_{j})^2$.

In order to solve for the emitter dynamics and the scattering of such a system in a waveguide, we invoke the photon-scattering relation of Eq. (\ref{eq5}). As part of the effective operator method \cite{Reiter12}, we can write the Hamiltonian in standard notation according to Eq. (\ref{eq17}). In Eq. (\ref{eq17a}), $\tilde{\Delta}_{e} = \Delta_{e}-E_g/\hbar$, with $\Delta_{e} = \mathcal{H}_{0}-\omega$. Here $\omega$ is the central frequency of the incoming light field and $E_{g}$ is the energy of the ground-state we excite out from. From here we see that the non-Hermitian Hamiltonian is initial-state (ini) dependent. When writing  $\hat{\mathcal{H}}_{\text{nh}}=\hat{\mathcal{H}}_e^{(\text{ini})}-\frac{i}{2}\sum_{k}\hat{\mathcal{L}}^\dagger_{k}\hat{\mathcal{L}}_{k}$ with 
\bea
\label{lindbladLambda}
\hat{\mathcal{L}}_{1,j} & = &\hat{\mathcal{L}}^{'}_{j} = \sqrt{\Gamma'_j}~\hat{\sigma}_{2j},\\
\label{lindbladLambda1}
\hat{\mathcal{L}}_{2,j} & = &\hat{\mathcal{L}}^{R}_{j} = \sqrt{\Gamma^{R}_{j,1\text{D}}}~\hat{\sigma}_{2j},\\
\label{lindbladLambda2}
\hat{\mathcal{L}}_{3,j} & = &\hat{\mathcal{L}}^{L}_{j} = \sqrt{\Gamma^{L}_{j,1\text{D}}}~\hat{\sigma}_{2j},
\eea 
two initial-state dependent Hamiltonians emerge:
\begin{subequations}\label{Hnhlambda}
\begin{equation}
\hat{\mathcal{H}}_{\text{nh}}^{(0)}=\left(\delta_0-\frac{i\Gamma}{2}\right)\hat{\sigma}_{22}\equiv\tilde{\delta}_0\hat{\sigma}_{22},
\end{equation}
\begin{equation}
\hat{\mathcal{H}}_{\text{nh}}^{(1)}=\left(\delta_1-\frac{i\Gamma}{2}\right)\hat{\sigma}_{22}\equiv\tilde{\delta}_1\hat{\sigma}_{22},
\end{equation}
\end{subequations}
which describe the excited-subspace energies and decay rates corresponding to excitation out of the two different ground-states. Here, we have changed to a rotating frame where $\delta_0 = (\omega_{22}-\omega_{00}-\omega)$ and $\delta_1=(\omega_{22}-\omega_{11}-\omega)$. The total decay rate of the excited state $|2\rangle$ is defined as $\Gamma=\Gamma_{0,1\text{D}}+\Gamma_{1,1\text{D}}+\Gamma_{0}'+\Gamma_1'$, where $\Gamma_{j,1\text{D}} = \sum_\zeta\Gamma^{\zeta}_{j, 1\text{D}}$ is the total decay rate for all transitions out of $|2\rangle$ into the state $|j\rangle$ by emitting into the $1\text{D}$ waveguide mode. 

Now, let us assume that the energy separation between the ground-states is much larger than the linewidths of all states, such that the incoming field only drives a single transition. We pick the exciting transition to be from $|0\rangle$ to $|2\rangle$, which can subsequently decay to either ground-state. From here on, we therefore omit the indices on $\hat{\mathcal{H}}_{\text{nh}}^{(j)}$ and $\delta_j$. Inverting the non-Hermitian Hamiltonian in Eq. (\ref{Hnhlambda}) is straightforward and yields
\begin{equation}\label{HNHlambda}
\hat{\mathcal{H}}_{\text{nh}}^{-1}=\tilde{\delta}^{-1}\hat{\sigma}_{22}
\end{equation}
where $\tilde{\delta}^{-1}\equiv (\delta-\frac{i}{2}\Gamma)^{-1}$. For an incoming photon incident from the left end of the double-sided waveguide and travelling towards the right, we can write Eq. (\ref{eq5}) in terms of the electric field on the left and right after scattering from the $\Lambda$-system emitter as:
\bea
\label{eq73}
&&\hat{a}_{out,\text{R}}(\text{z},t) = \Bigg[1+i\Bigg(\Gamma^{\text{R}}_{0,1\text{D}}\tilde{\delta}^{-1}\hat{\sigma}_{00}+\tilde{\delta}^{-1}\sqrt{\Gamma^{\text{R}}_{0,1\text{D}}\Gamma^{\text{R}}_{1,1\text{D}}}\nonumber\\
&&\times\hat{\sigma}_{01}e^{-i\omega_{01}\frac{(\text{z}-\text{z}_0)}{v_g}}\Bigg)\Bigg]\hat{a}_{in,\text{R}}(\text{z}-v_g t),\\
\label{eq74}
&&\hat{a}_{out,\text{L}}(\text{z}',t)= i\Bigg(\sqrt{\Gamma^{\text{L}}_{0,1\text{D}}\Gamma^{\text{R}}_{0,1\text{D}}}\tilde{\delta}^{-1}\hat{\sigma}_{00}+\tilde{\delta}^{-1}\sqrt{\Gamma^{\text{L}}_{1,1\text{D}}}\nonumber\\
&&\times\sqrt{\Gamma^{\text{R}}_{0,1\text{D}}}\hat{\sigma}_{01}e^{-i\omega_{01}\frac{(\text{z}_0-\text{z}')}{v_g}}\Bigg)e^{2ik_0(\text{z}_0-\text{z}')}\hat{a}_{in,\text{R}}(\text{z}'+v_g t),~
\eea
where $\text{z}_0$ is the position of the emitter, $\hat{\sigma}_{00} = |0\rangle\langle 0|$ and $\hat{\sigma}_{01} = |1\rangle\langle 0|$, while $\text{z}$ is some point to the right of the emitter and $\text{z}' < \text{z}_0$ is to the left of the emitter.

From Eqs. (\ref{eq73}) and (\ref{eq74}) we see that, unlike the earlier discussed cases involving only the population of a single ground-state, the scattered field now involves the response of the emitter in terms of both the population and coherence of the ground-states. Furthermore, compared to the previous examples now the populations of the ground-states $|0\rangle$ and $|1\rangle$ evolve with time. Hence we now need to invoke the effective-operator master equation (\ref{eq6a}) to solve for the dynamics of the emitter. To use the master equation we first define a basis $\{|0\rangle,|1\rangle\}$ with $\hat{\sigma}_{ij}=|j\rangle\langle i|$. The \textit{effective} Hamiltonian governing the coherent dynamics of the ground-state density matrix is given by
\begin{equation}\label{Heffref}
\hat{\mathcal{H}}_{\text{eff}}=-\frac{1}{2}\hat{\mathcal{V}}_- [\hat{\mathcal{H}}_{\text{nh}}^{-1}+(\hat{\mathcal{H}}_{\text{nh}}^{-1})^\dagger]\hat{\mathcal{V}}_{+}+\hat{\mathcal{H}}_g,
\end{equation}
where the excitation and de-excitation operators are defined respectively by $\hat{\mathcal{V}}_{+}=\sum_\mu\mathcal{A}^{\mu}_{0} \hat{a}_\mu \hat{\sigma}_{02}+\sum_\mu\mathcal{A}^{\mu}_{1} \hat{a}^{'}_\mu \hat{\sigma}_{12}$ and $\hat{\mathcal{V}}_{-} = \sum_\mu\mathcal{A}^{\mu}_{0} \hat{a}^\dagger_\mu \hat{\sigma}_{20}+\sum_\mu\mathcal{A}^{\mu}_{1} \hat{a}^{'\dagger}_\mu \hat{\sigma}_{21}$ while  $\hat{\mathcal{H}}_g =\omega_{01}\hat{\sigma}_{11}$. Here the prime on the mode operator reflects that the field needs to have different frequencies to be resonant with the two different transition. As in this work we are mainly interested in the regime where the splitting between the ground states is large compared to the optical line width, the corresponding mode operators can essentially be considered to represent two different baths. Recall that $ \mathcal{A}^{\zeta}_{j,(1\text{D})}=\sqrt{\Gamma_{j,1\text{D}}^{\zeta}}$. Note that as opposed to the previous examples we will here need to be careful about the noise terms in the $\hat{\mathcal{H}}_{\text{eff}}$. Such noise terms arise due to contribution from modes outside of the waveguide in $\hat{\mathcal{V}}_{\pm}$. Using the above expressions for $\hat{\mathcal{V}}_{\pm}$ and Eq. (\ref{HNHlambda}) we then evaluate $\mathcal{H}_{\text{eff}}$ to be
\begin{equation}
\mathcal{H}_{\text{eff}} = \left[\begin{array}{cc} - (\sum_{\zeta\zeta'}\sqrt{\Gamma^{\zeta}_{0,1\text{D}}}\sqrt{\Gamma^{\zeta'}_{0,1\text{D}}}\hat{a}_{\zeta}^\dagger \hat{a}_{\zeta'})\frac{\delta}{|\tilde{\delta}|^2}+\mathcal{F}  & \mathcal{F}^{'}  \\
\\
 \mathcal{F}^{'\dagger} & \omega_{01}
\end{array}\right].
\end{equation}
Here the noise terms $\mathcal{F}$ and $\mathcal{F}'$ are given respectively by $\mathcal{F} = -[\sum_{\zeta}\sqrt{\Gamma^{\zeta}_{0,1\text{D}}}\sqrt{\Gamma^{'}_{0}}~\hat{a}_{\zeta}^\dagger \hat{a}_{s}+\sum_{\zeta}\sqrt{\Gamma^{\zeta}_{0,1\text{D}}}\sqrt{\Gamma^{'}_{0}}~\hat{a}_{s}^\dagger \hat{a}_{\zeta} +\Gamma^{'}_{0}~\hat{a}_{s}^\dagger \hat{a}_{s}] \delta/|\tilde{\delta}|^2$ and $\mathcal{F}' = -\sum_{\mu}\sum_{\mu'}\mathcal{A}^{\mu}_{0}\mathcal{A}^{\mu'}_{1}\hat{a}^{\dagger}_{\mu}\hat{a}^{'}_{\mu'}(\delta/|\tilde{\delta}|^2)$. Furthermore, in writing the $|1\rangle\langle 1|$ element of the matrix $\mathcal{H}_{\text{eff}}$, we have neglected the terms $\sum_{\mu}\sum_{\mu'}\mathcal{A}^{\mu}_{1}\mathcal{A}^{\mu'}_{1}\hat{a}^{'\dagger}_{\mu}\hat{a}^{'}_{\mu'}$. This is because there are no photons at the frequency corresponding to the primed reservoir since we assume that the incoming field is resonant with the transition $|0\rangle\rightarrow|2\rangle$. Also, we define \textit{effective} Lindblad decay operators in the form
\begin{equation}\label{LKchannels}
\hat{\mathcal{L}}^k_{\text{eff}}=\hat{\mathcal{L}}_k \hat{\mathcal{H}}_{\text{nh}}^{-1}\hat{\mathcal{V}}_{+},
\end{equation} 
for each decay channel $k$. Recall that as $\hat{\mathcal{V}}_{\pm}$ includes modes outside the waveguide, $\hat{\mathcal{L}}^k_{\text{eff}}$ also has contribution from the noise in the system dynamics. In the $\Lambda$-system, we drive only the transition from $|0\rangle$ to $|2\rangle$, which can decay to either $|0\rangle$ or $|1\rangle$. We then only have two effective decoherence channels: \textit{population transfer} described by $|1\rangle \langle 0|$ and a driving-induced \textit{dephasing} term (shift) described by $|0 \rangle \langle 0|$. Plugging Eq. (\ref{Vpequation}), Eqs. (\ref{lindbladLambda})- (\ref{lindbladLambda2}), and Eq. (\ref{HNHlambda}) into Eq. (\ref{LKchannels}), we find the following effective Lindblad operators:
\begin{eqnarray}\label{LchannelsEff}
\hat{\mathcal{L}}'_{\text{eff}} & = &\tilde{\delta}^{-1}\sum^{1}_{j = 0}\sqrt{\Gamma_{j}'}\sum_{\zeta'}\sqrt{\Gamma^{\zeta'}_{0, 1\text{D}}}~\hat{\sigma}_{0j}\hat{a}_{\zeta'},\nonumber\\
&+&\tilde{\delta}^{-1}\sum^{1}_{j = 0}\sqrt{\Gamma_{j}'}\sqrt{\Gamma^{'}_{0}}~\hat{\sigma}_{0j}\hat{a}_{s},\\
\hat{\mathcal{L}}^{\zeta}_{\text{eff}} & = &\tilde{\delta}^{-1}\sum^{1}_{j = 0}\sqrt{\Gamma^\zeta_{j, 1\text{D}}}\sum_{\zeta'}\sqrt{\Gamma^{\zeta'}_{0, (1\text{D})}} \hat{\sigma}_{0j}\hat{a}_{\zeta'}\nonumber\\
& + &\tilde{\delta}^{-1}\sum^{1}_{j = 0}\sqrt{\Gamma^\zeta_{j, 1\text{D}}}\sqrt{\Gamma^{'}_{0}}~\hat{\sigma}_{0j}\hat{a}_{s}.
\end{eqnarray}
We next assume that the coupling of the photon to the right and left travelling mode in the waveguide have the same strength such that $\Gamma^{\text{R}}_{j,1\text{D}} = \Gamma^{\text{L}}_{j,1\text{D}} = \Gamma_{j,1\text{D}}/2$. Also, we consider the incoming field only to be only in the right-propagating mode, such that $\hat{a}_{\text{L},in}|\Psi_{ini}\rangle = 0$. Hence, for all further discussions the scattered field-mode will depend only on $\hat{a}_{\text{R}, in}$ with the other modes $\hat{a}_{s}$ and $\hat{a}_{\text{L}}$ contributing to the losses and noise. For notational simplicity we will represent $\hat{a}_{\text{R}, in}$ by $\hat{a}$, while all terms containing $\hat{a}_{s}$ and $\hat{a}_{\text{L}}$  will be called noise.

Combining the above considerations with Eq. (\ref{Heffref}) and Eq. (\ref{LchannelsEff}), we evaluate the effective master equation Eq. (\ref{eq6a}) for each element in the ground-state density matrix. This gives a series of coupled-component differential equations,
\begin{subequations}\label{masterEqDiffs}
\begin{align}
\dot{\hat{\sigma}}_{00} &= \colon-\text{P}_{R}\hat{a}^\dagger\hat{a}\hat{\sigma}_{00}\colon + \text{Noise}\\
\dot{\hat{\sigma}}_{11} &= \colon+\text{P}_{R}\hat{a}^\dagger\hat{a}\hat{\sigma}_{00}\colon+\text{Noise},\\
\dot{\hat{\sigma}}_{01} &= \colon +i \hat{\sigma}_{01}(\mathcal{H}_{\text{eff},22}-\mathcal{H}_{\text{eff},11})\nonumber\\ &-\frac{1}{2}\left(\text{P}_{R}+\text{P}_{d}\right)\hat{a}^\dagger\hat{a}\hat{\sigma}_{01}\colon+\text{Noise},\\
\dot{\hat{\sigma}}_{10} &= \colon-i \hat{\sigma}_{10}(\mathcal{H}_{\text{eff},22}-\mathcal{H}_{\text{eff},11})\nonumber\\ &-\frac{1}{2}\left(\text{P}_{R}+\text{P}_{d}\right)\hat{a}^\dagger\hat{a}\hat{\sigma}_{10}\colon+\text{Noise},
\end{align}
\end{subequations}
where $\mathcal{H}_{\text{eff},jj} = \langle j|\mathcal{H}_{\text{eff}}|j\rangle$ in Eq. (\ref{Heffref}) and the effective \textit{probabilities} corresponding to the amplitudes of the operators in the above equations. These are given by 
\bea
\label{prob1}
\text{P}_{d}& = &\frac{\Gamma_{0}\Gamma^R_{0,1\text{D}}}{|\tilde{\delta}|^2},\\
\label{prob2}
\text{P}_{R}&=&\frac{\Gamma_{1}\Gamma^R_{0,1\text{D}}}{|\tilde{\delta}|^2},
\eea
where $\text{P}_{d}$ represents the photon induced dephasing of level $|0\rangle$ while $\text{P}_{R}$ represents the total Raman scattering probability, i.e., the probability for a single photon to scatter $|0\rangle\rightarrow|2\rangle\rightarrow|1\rangle$, either emitting into the waveguide in either direction, or to the side. To find these probabilities we have evaluated quantities like $\text{P}_{d}\hat{a}^\dagger \hat{a} = \sum_{k = ', \text{R},\text{L}}\langle 0 |\hat{\mathcal{L}}^{k\dagger}_{\text{eff}} |0\rangle\langle 0 |\hat{\mathcal{L}}^{k}_{\text{eff}} | 0 \rangle$ and $\text{P}_{R} \hat{a}^\dagger \hat{a} = \sum_{k = ', \text{R},\text{L}}\langle 0 | \hat{\mathcal{L}}_{\text{eff}}^{k\dagger} |1\rangle  \langle 1 | \hat{\mathcal{L}}^{k}_{\text{eff}} | 0 \rangle$.

The solution of the above set of equations is straightforward. In particular, we find the solution of the ground-state occupations to be
\begin{subequations}
\begin{align}
\label{grnd1}
\hat{\sigma}_{00}(t)&=\colon\hat{\sigma}_{00}(0) e^{-\text{P}_{R}\int_{0}^{t}\hat{a}^\dagger\hat{a} dt'}\colon + \text{Noise},\\
\label{grnd2}
\hat{\sigma}_{11}(t)&=\colon(1-\hat{\sigma}_{00}(0)e^{-\text{P}_{R}\int_{0}^{t}\hat{a}^\dagger\hat{a} dt'})\colon + \text{Noise}.
\end{align}
\end{subequations}
Thus we see from the solution of the master equation that the input field drives the population from $|0\rangle$ to $|1\rangle$ at a rate $\text{P}_{R} \hat{a}^\dagger \hat{a}$, that is proportional to the input-field operators appearing in the excitation terms $\hat{\mathcal{V}}_{+}$ in the effective decay channels $\hat{\mathcal{L}}^{k}_{\text{eff}}$ in Eq. (\ref{LchannelsEff}). 
\subsection{The Photon Scattering Dynamics}
Now that we have the knowledge of all the relevant dynamics, let us investigate light scattering into the waveguide from the emitter. To elucidate the scattering problem further, we in the following subsections consider three specific cases: (1) single-photon scattering and the probability of photo-detection after separating the two frequency components in the scattered field via a filter, (2) coherent pulse scattering followed by intensity measurement of unfiltered output, and lastly (3) generation of a ground-state superposition conditioned on photodetection (click of the detector). For all the cases discussed below, we assume that the coupling to both the right-propagating and the left-propagating modes in the waveguide are equal i.e., $\Gamma^{R} _{j,1\text{D}} = \Gamma^{L} _{j,1\text{D}} = \Gamma_{j,1\text{D}}/2$.

\subsubsection{Frequency filtering of scattered single photon}
Let us assume that the input field has a single near resonant photon only. The photon can excite the $|0\rangle$ to $|2\rangle$ transition, and a photon comes out either at the input photon frequency $\omega = (\omega_{22}-\omega_{00})-\delta_{0}$ (blue) or at $\omega_{12}=(\omega_{22}-\omega_{11})-\delta_{1}$ (red). In labelling the photon as red and blue we have assumed $\omega_{11} > \omega_{00}$. If the emitter starts in one ground-state, the outgoing photon becomes entangled with the emitter ground-state $|0\rangle$ or $|1\rangle$. By removing for example blue photons from the output using a filter, we can condition the experiment on a click in a detector to say that the emitter has flipped from state $|0\rangle$ to $|1\rangle$. Mathematically, the frequency shift is, in our formalism, contained in the time evolution of the $\hat{\sigma}_{01}$ operator in Eq. (\ref{eq73}) and Eq. (\ref{eq74}). The action of the frequency filter thus amounts to only retaining the term containing $\hat{\sigma}_{01}$ in Eqs. (\ref{eq73}) and (\ref{eq74}). We name the filtered $\hat{a}_{out,\text{R}}$ as $\hat{a}_{out,\text{R}, red}$ and henceforth use it to denote the filtered output. 

If we consider a single right-going photon input, the probability of getting a right-going red photon coming out is given by
\bea
\label{eq75}
\text{P}^{R}_{red}&\sim&\frac{\int\langle\Psi_{ini}|\hat{a}^\dagger_{out,\text{R},red}(t)\hat{a}_{out,\text{R},red}(t)|\Psi_{ini}\rangle dt}{\int\langle\Psi_{ini}|\hat{a}^\dagger_{in,\text{R}}(t)\hat{a}_{in,\text{R}}(t)|\Psi_{ini}\rangle dt}\nonumber\\
& = & \left|\frac{\sqrt{\Gamma^{\text{R}}_{0,1\text{D}}\Gamma^{\text{R}}_{1,1\text{D}}}}{\tilde{\delta}}\right|^{2}\nonumber\\
&\times&\int\langle\Psi_{ini}|\hat{a}_{in,\text{R}}^\dagger\hat{\sigma}_{01}(t)\hat{\sigma}_{10}(t)\hat{a}_{in,\text{R}}|\Psi_{ini}\rangle dt,
\eea
where, $|\Psi_{ini}\rangle \equiv \hat{a}^\dagger_0|0,\varnothing\rangle$ is the initial state of the total system with the emitter in state $|0\rangle$ and incoming right-going single-photon creation operator $\hat{a}_{0}^{\dagger}=\int dk F_{R,k}^\dagger \hat{a}_k^\dagger $, for some suitable mode function $F_{R,k}$ such that $\int \langle \Psi_{ini}|\hat{a}_{in,\text{R}}^\dagger(t)\hat{a}_{in,\text{R}}(t)|\Psi_{ini}\rangle dt =1$. Using $e^{x}=\sum_{k=0}^{\infty} x^k/k!$ and normal ordering the solution in Eq. (\ref{grnd1}), the evaluation of the integral $\int\langle\Psi_{ini}|\hat{a}_{in,\text{R}}^\dagger\hat{\sigma}_{01}(t)\hat{\sigma}_{10}(t)\hat{a}_{in,\text{R}}|\Psi_{ini}\rangle dt$ yields $\langle 0,\emptyset_{R}|\hat{\sigma}_{00}(0)| 0,\emptyset_{R}\rangle$, where we have used that all noise operators vanish for a vacuum input state.

Now, as the $\Lambda$-system is assumed to be initially prepared in the ground-state $|0\rangle$, we have $\langle \hat{\sigma}_{00} (0)\rangle = 1$. Thus, on substituting this in Eq. (\ref{eq75}) we find 
\bea
\label{eq76}
\text{P}^{R}_{red}= \frac{\beta_{0}\beta_{1}}{\left(1+\frac{4\delta^2}{\Gamma^2}\right)}~, 
\eea
where $\Gamma=\Gamma_{0,1\text{D}}+\Gamma_{1,1\text{D}}+\Gamma_{0}'+\Gamma_1'$ is the total decay rate while $\beta_{0} = \Gamma_{0,1\text{D}}/\Gamma$ and $\beta_{1} = \Gamma_{1,1\text{D}}/\Gamma$. As we assumed equal rates of decay to the left and right, $\text{P}^{R}_{red}=\text{P}^{L}_{red}$ and the scattering probability is maximal for $\Gamma_{0,1\text{D}} = \Gamma_{1,1\text{D}}$ with $\Gamma'=0$ and on resonance $\delta=0$. For these parameters, a single photon has a $50\%$ chance to flip the emitter, and a red photon is emitted left or right with equal probabilities to yield a total probability of $25\%$ for detecting the photon. Note that here the normal ordering of the operators in Eqs. (\ref{grnd1}) and (\ref{grnd2}) is essential for getting the right results. Without normal ordering the result in Eq. (\ref{eq76}) would contain higher-order terms in the probability, which should not be there for a single incident photon. Likewise, we can perform filtered detection of a \textit{blue} photon, yielding $\text{P}^{R}_{blue} = 1- (2-\beta_{0})\beta_{0}/\left(1+\frac{4\delta^2}{\Gamma^{2}}\right)$.

\subsubsection{Unfiltered total intensity output for a coherent pulse input}
Instead of a single photon, if we use a weak coherent pulse as an input field, the scattering dynamics is different. In this situation, a coherent pulse input can drive the emitter from the ground-state $|0\rangle$ to $|1\rangle$ before the detection time that we consider, since now the incoming pulse may contain more than one photon. To study the characteristic of the transmitted field, we again use Eq. (\ref{eq73}). Typically, in experiments one measures the intensity of the output field using photo-detectors, so we calculate the expectation value of the square of the output-field operator (without any filtering) as
 \begin{equation}\label{outputIntensityLambda}
I_{out}=\langle \hat{a}_{out}^\dagger \hat{a}_{out} \rangle = \langle \Psi_{ini} | \hat{a}_{out}^\dagger \hat{a}_{out} | \Psi_{ini} \rangle,
\end{equation}
where $|\Psi_{ini}\rangle$ is the initial state of the emitter-field system. If we as before choose the emitter to be prepared initially in the state $|0\rangle$ while the field is in the coherent state $|\alpha\rangle$ such that $| \Psi_{ini} \rangle = | \Psi_{\alpha}\rangle =|0,\alpha\rangle $, we get the intensity
\begin{multline}
I_{out}=\langle\Psi_{\alpha}| \hat{a}^\dagger \Big[1-\frac{(2-\beta_{0}-\beta_{1})\beta_{0}}{\left(1+\frac{4\delta^2}{\Gamma^2}\right)}\hat{\sigma}_{00}(t)\Big] \hat{a} | \Psi_{\alpha}\rangle
\end{multline}
where $\Gamma$ is the total decay rate of the excited level, $\Gamma=\Gamma_{0,1\text{D}}+\Gamma_{1,1\text{D}}+\Gamma_{0}'+\Gamma_1'$. In this calculation we evaluate the time-dependent density matrix element $|0\rangle\langle 0|$ decaying with the probability $\text{P}_R=\frac{\Gamma_{1}\Gamma_{0,1\text{D}}}{2|\tilde{\delta}|^2}$ per incident photon.

Let us now evaluate the term $\langle\Psi_{\alpha}| \hat{a}^\dagger \hat{\sigma}_{00}(t) \hat{a} | \Psi_{\alpha}\rangle$. Note that $\hat{a}^\dagger \hat{a}$ is in the exponential in the solution given in Eq. (\ref{grnd1}) which in turn can be written as a power series $e^{x}=\sum_{k=0}^{\infty} x^k/k!$. Also, recall that the solution to the master equation assumes normal ordering of the field-mode operators, such that $\langle \alpha |\colon\hat{a}^\dagger(\sum_{k=0}^{\infty} (\hat{a}^\dagger \hat{a})^k/k!) \hat{a}\colon| \alpha \rangle = \sum_{k=1}^{\infty} (\alpha^*\alpha)^k/(k-1)!$. Using this we then get,  $\langle \alpha |\colon\hat{a}^{\dagger}e^{-\text{P}_R \int_{0}^{t}\hat{a}^\dagger\hat{a} dt'} \hat{a}\colon|\alpha\rangle = |\alpha(t)|^{2}e^{-\text{P}_R |\alpha(t)|^2 t}$, where $|\alpha(t)|^2$ is the intensity of the coherent state $|\Psi_\alpha\rangle$. Also, as before we then choose the initial state such that $\langle 0|\hat{\sigma}_{00}(t=0)|0\rangle=1$. We can then write
\begin{equation}
\label{intensityRef}
I_{out}(t)=|\alpha(t)|^2 \left(1 - \text{P}_{sc} e^{-\text{P}_R \int_{0}^{t}|\alpha(t')|^2 dt'} \right), 
\end{equation}
where the time $t=0$ is defined as the moment the incident pulse reaches the emitter, and 
\bea
\label{psuc}
\text{P}_{sc} = \frac{(2-\beta_{0}-\beta_{1})\beta_{0}}{\left(1+\frac{4\delta^2}{\Gamma^{2}}\right)} 
\eea
is the probability for a single photon to scatter into other directions than the right-going guided mode.  

Let us now consider the probability of a click (photo-detection event) at a detector placed to the right of the emitter. If the input was a single photon, the probability of detecting a (any colour) right going photon would be
\begin{equation}
\text{P}_{click}^{(1)} = \eta \left(1 - P_{sc} \right) =\eta (\text{P}^{R}_{red}+\text{P}^{R}_{blue}).
\end{equation} 
This, e.g., reduces to $\eta$, the detection efficiency, for $\Gamma_{0,1\text{D}}=0$, where there is no interaction with the emitter, and goes to zero for $\Gamma_{1,1\text{D}} = \Gamma'=\delta=0$ which is a perfectly reflecting two-level system. If we have a resonant field with no decay to the side, $\Gamma'=\delta=0$, and equal decay rates $\Gamma_{0,1\text{D}} = \Gamma_{1,1\text{D}} = \Gamma_{1\text{D}}$, there will be a $50\%$ chance of passing through to the right.

If, instead, the input is a weak coherent pulse, we need to integrate the output intensity over the pulse duration $T$ of the input to find the total number of photons in the output. We consider a weak pulse, such that the integration yields the \textit{probability} of detecting even a single photon. For a coherent pulse of duration $T$, we can define a total input photon number $\bar{n}=\int_0^T |\alpha(t)|^2 dt$. Thus, using Eq. (\ref{intensityRef}) we get the detection probability for $\text{P}_{click}^{(c)}\ll 1$ as 
\begin{subequations}
\begin{align}
\text{P}_{click}^{(c)}&=\eta \int_0^T I_{out}(t) dt\nonumber\\&=\eta\left[\bar{n}-\frac{\text{P}_{sc}}{\text{P}_{R}}\left[1-e^{-\text{P}_{R}\bar{n}}\right]\right]\nonumber\\
&\approx \eta \bar{n} (1-\text{P}_{sc}) =\bar{n} \text{P}_{click}^{(1)},
\end{align}
\end{subequations}
where the last approximation is valid in the limit $\text{P}_R\bar{n}\ll1$; In this limit, the number of detected photons is to first order proportional to $\text{P}^{(1)}_{click}$, the probability of transmitting a single photon to the right.

\subsubsection{Conditional generation of ground-state superposition }
In this example, we demonstrate how our formalism can be used to describe conditional state preparation in a $\Lambda$ type emitter.  In particular, our objective is to create a superposition state of the emitter's ground levels of the form $|\Psi^{-}\rangle=(|0\rangle-|1\rangle)/\sqrt{2}$. The physics of this state creation process is as follows. Due to the two transition pathways in a $\Lambda$ system, a photon-scattering process leads to an entangled state of light and matter of the form $|\Psi_{\text{ent}}\rangle = \frac{1}{\sqrt{2}}\left(|\omega_{\text{blue}}\rangle|0\rangle-|\omega_{\text{red}}\rangle|1\rangle\right)$, where $(\omega_{\text{blue}}-\omega_{\text{red}}) = \omega_{01}$, and where $|\omega\rangle$ refers to a single photon state with frequency $\omega$. Without filtering, the frequency difference between the two ground-states encoded in the outgoing photon will remain unresolved. A click in the photo-detector at a certain time $t$ will erase the `which path' information of the scattering, thereby creating the superposition state $|\Psi^{-}\rangle$.

Let us next evaluate the fidelity of being in state $|\Psi^{-}\rangle=(|0\rangle-|1\rangle)/\sqrt{2}$:
\begin{equation}
\label{fidelity}
F=\langle \Psi^{-}| \hat{\rho}^{(c)} | \Psi^{-}\rangle = \frac{1}{2} (\rho_{00}^{(c)}-\rho_{01}^{(c)}-\rho_{10}^{(c)}+\rho_{11}^{(c)}),
\end{equation}
where the elements $\rho_{ij}^{(c)}$ of the conditional density matrix $\rho^{(c)}$ can be evaluated from Eq. (\ref{generalsigma}) below. Note that, due to normalisation, $\rho_{11}^{(c)}+\rho_{00}^{(c)} = \text{Tr}(\hat{\rho}^{(c)}) = 1$ and we only need to evaluate the coherence $\rho^{(c)}_{01}$.

We next lay down a mathematical treatment for the state creation process. We begin by considering the evolution of the density matrix elements under the influence of an incoming coherent pulse. Recall that the output-field operator is also a function of the emitter operators. To find the total system evolution, we write the density matrix conditioned on a click in a detector at time $t_{c}$
\begin{equation}\label{generalsigma}
\rho^{(c)}_{ij} (t_c, T) =\frac{\langle \Psi_{ini}|\hat{a}^\dagger_{out}(t_c)\hat{\sigma}_{ij}(T)\hat{a}_{out}(t_c)|\Psi_{ini}\rangle}{\langle \Psi_{ini}|\hat{a}^\dagger_{out}(t_c)\hat{a}_{out}(t_c)|\Psi_{ini}\rangle}.
\end{equation}
In Eq. (\ref{generalsigma}), we condition on having a click at a certain time $t_{c}$, represented by the operators $\hat{a}_{out}$. Experimentally one would however, only consider the first click which arrives at the detector. This makes no difference if the incident field only contains a single photon since in this case one cannot have two clicks. With an incident coherent state a more correct description would be to include in Eq. (\ref{generalsigma}) the requirement that there is no photon detected before the time $t_{c}$. Since we mainly consider the limit where the probability of a detection event is small, the probability of having two detection events in the time interval is negligible and the simple description in Eq. (\ref{generalsigma}) is sufficient. Furthermore, we wish to calculate the time evolution of $\rho_{01}^{(c)}$ until a point $T$, i.e., the full duration of the incoming pulse sequence. After that, we know that the free evolution of the coherence will simply oscillate with the energy difference between the ground-states. Recall that $t_{c}$ is the time after the start of the pulse, at which a photon was detected by click in the photo-detector and hence in this experiment we have $t_{c} \leq T$.

In evaluating Eq. (\ref{generalsigma}) we have to be extra careful as now the vacuum noise operators, which until now we have neglected play a crucial role in the dynamics of $\rho^{(c)}_{ij}$.  In particular for coherence term like $\rho^{(c)}_{01}$, one has to evaluate quantities like $\hat{a}^\dagger_{out}(t_c)\sigma_{01}(T)\hat{a}_{out}(t_c)$. From Eq. (\ref{eq73}) and Eq. (\ref{masterEqDiffs}) we see that this will then involve terms like $\hat{\sigma}_{01}(t_{c})\hat{\sigma}_{01}(T)\hat{\sigma}_{00}(t_{c})$. Here we need to evaluate a product of operators at different times. With the normal ordered operators from in Eq. (\ref{grnd1}) we have ensured that the noise operators for each of the terms vanish. This is, however, no longer the case once we have the product of three normal ordered terms and in principle we need to evaluate the noise terms. To avoid this complication we instead first calculate $\rho^{(c)}(t_c,t_c)$. In this case the three operators obey the relation $\hat{\sigma}_{10}(t_c)\hat{\sigma}_{01}(t_{c})\hat{\sigma}_{00}(t_c) = \hat{\sigma}_{00}(t_c)$ since now all time arguments are equal (recall here the definition $\hat{\sigma}_{ij} = |j\rangle\langle i|$, which leads to unconventional rules for the indices in products of operators). With this relation we have reduced the product of three operators to a single operator. We can then simply use Eq. (\ref{grnd1}) for a single time and all noise operators are normal ordered such that they vanish for initial vacuum states. To find the final density matrix $\rho^{(c)}(t_c,T)$, we then evolve the density matrix $\rho^{(c)}$ from $t_c$ to $T$. Using  Eq. (\ref{masterEqDiffs}) this gives us 
\begin{equation}
\label{eq81}
\rho_{01}^{(c)}(t_{c},T)=\rho_{01}^{(c)}(t_{c},t_{c}) e^{\int_{t_c}^{T}i \omega_{01}'-\frac{1}{2}(\text{P}_{R}+\text{P}_{d})|\alpha(t)|^2 dt},
\end{equation}
which essentially says that the coherence decays at a rate $\frac{1}{2}(\text{P}_{R}+\text{P}_{d})|\alpha(t)|^2$ over a time $(T-t_{c})$, due to both the Raman transfer rate and the photon-induced dephasing rate. Also, its phase rotates at a frequency $\omega_{01}'$ equal to the splitting between the two ground-states $|0\rangle$ and $|1\rangle$, $\omega_{01}$, plus some AC-Stark shift $\delta\omega= (\omega_{01}'-\omega_{01})$ induced by the weak coherent drive of the $|0\rangle$ ground-state, given by $\delta\omega = \langle\mathcal{H}^{\text{eff}}_{11}\rangle=\Gamma_{0,1\text{D}} |\alpha(t)|^2\delta/|\tilde{\delta}|^2$. 

Now we find the time evolution from $t = 0$ to the time of the click $t_{c}$ at the detector. Inserting the output field $\hat{a}_{out}$ in Eq. (\ref{generalsigma}) yields the elements as follows:
\begin{align}
\label{sigmac}
\rho_{01}^{(c)}(t_{c},t_{c})&=\frac{\langle \Psi_{ini}|\hat{a}^\dagger_{out}(t_c)\hat{\sigma}_{01}(t_{c})\hat{a}_{out}(t_c)|\Psi_{ini}\rangle}{\langle \Psi_{ini}|\hat{a}^\dagger_{out}\hat{a}_{out}|\Psi_{ini}\rangle}.
\end{align}
The denominator of Eq (\ref{sigmac}), can be recognized as the intensity of the output, given by $I_{out}(t)=|\alpha(t)|^2 \left(1 - \text{P}_{sc} e^{-\text{P}_R \int_{0}^{t}|\alpha(t')|^2 dt'} \right)  $.
\begin{figure}
\begin{tabular}{ccc}
\includegraphics[width=0.24\textwidth]{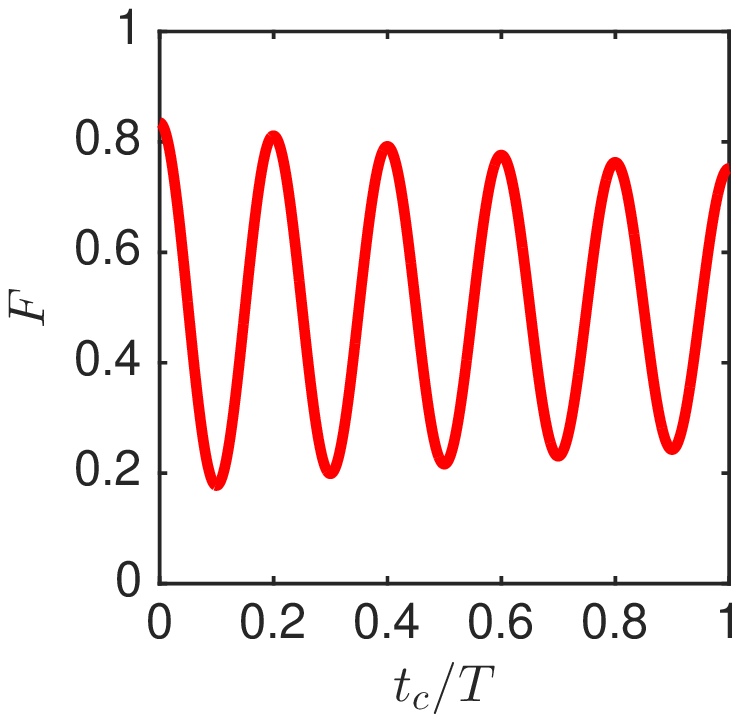} &  \includegraphics[width=0.24\textwidth]{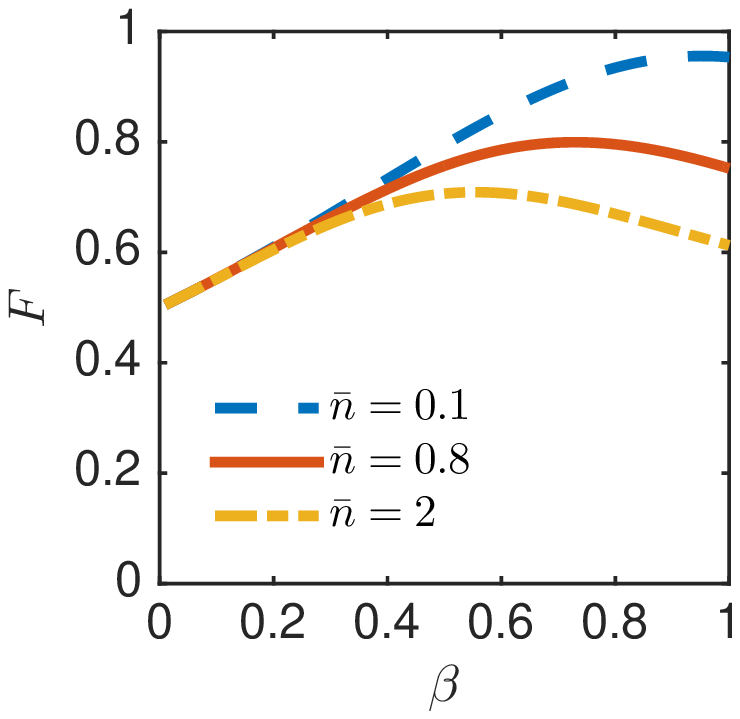}\\
(a) & (b) 
\end{tabular}
\caption{(a) Fidelity of the antisymmetric superposition state $|\Psi^{-}\rangle$ as a function of the detection time $t_{c}$ normalized with the pulse duration $T$. We plot here for $\Gamma_{0,1\text{D}}=\Gamma_{1,1\text{D}}$, $\delta=0$, $\phi_z=0$, $\beta=1$, $\omega_{01} = 5 \frac{2\pi}{T}$, and an average number of photons $\bar{n} = 0.8$. Resolving the detection time determines the phase of the generated state. The detection time has an arbitrary offset determined by the spatial position of the detectors. (b) Fidelity of superposition-state generation as a function of the $\beta$-factor for different values of $\bar{n}$, the average photon number in the coherent pulse.}
\label{fidplot}
\end{figure}

Next, for notational convenience, let us write the output field $\hat{a}_{out}$ in Eq. (\ref{eq73}) in the form
\begin{equation}
\label{eq80}
\hat{a}_{out}=\left[1+i\left(A\hat{\sigma}_{00}+B\hat{\sigma}_{01}\right)\right]\hat{a}_{in},
\end{equation}
where we define $A = \Gamma_{0,1\text{D}}/2\tilde{\delta}$ and $B = \sqrt{\Gamma_{0,1\text{D}}\Gamma_{1,1\text{D}}}/2\tilde{\delta}~\exp[-i\omega_{01}(\text{z}-\text{z}_0)/v_\text{R}] $. 
Substituting Eq. (\ref{eq80}) into Eq. (\ref{sigmac}) we then get
\bea
\label{eq82}
\rho_{01}^{(c)}(t_{c},t_{c})& = &\langle \Psi_{ini}|\hat{a}^\dagger_{in}\left[1-i\left(A^*\hat{\sigma}_{00}+B^*\hat{\sigma}_{10}\right)\right]\hat{\sigma}_{01}(t_{c})\nonumber\\
&\times&\Big[1+i\Big(A\hat{\sigma}_{00}+B\hat{\sigma}_{01}\Big)\Big]\hat{a}_{in}|\Psi_{ini}\rangle/I_{out}(t_{c}).\nonumber\\
\eea
Considering only the relevant terms in Eq. (\ref{eq82}) we get 
\bea
\label{eq83}
\rho_{01}^{(c)}(t_{c},t_{c})&=&|\alpha(t_c)|^2\langle \Psi_{ini}|\left[1-i\left(A^*\hat{\sigma}_{00}+B^*\hat{\sigma}_{10}\right)\right]\nonumber\\
&\times&\hat{\sigma}_{01}(t_{c})\Big[1+i\left(A\hat{\sigma}_{00}+B\hat{\sigma}_{01}\right)\Big]|\Psi_{ini}\rangle/I_{out}(t_c).\nonumber\\
\eea

Now evaluating the expectation values of the operators $\langle \Psi_{ini} |\hat{\sigma}_{10}(t_{c})\hat{\sigma}_{01}(t_{c})\hat{\sigma}_{00}(t_{c})|\Psi_{ini}\rangle=\langle \Psi_{ini}|\hat{\sigma}_{10}(t_{c})\hat{\sigma}_{01}(t_{c}) |\Psi_{ini}\rangle$,  we get $\langle \Psi_{ini}|\hat{\sigma}_{00}(t=0)\colon e^{-\text{P}_R\int_{0}^{t_c}\hat{a}^\dagger \hat{a} dt}\colon |\Psi_{ini}\rangle=e^{-\text{P}_R\int_{0}^{t_c}|\alpha(t)|^2dt}$. Inserting the solution for $\rho_{01}^{(c)}(t_{c},t_{c})$ into Eq. (\ref{eq81}) gives us
 \bea
 \label{eq84}
 \rho_{01}^{(c)}(t_{c},T) & = & |\alpha(t_c)|^2 (1+i A)(-i B^*)\nonumber\\
 &\times& (e^{-\gamma(t_c,T)+\int_{t_c}^{T}i \omega_{01}'(t)dt})/I_{out}(t_{c}),
 \eea
where for notational convenience we have introduced a total `coherence-decay' term
\begin{multline} \gamma(t_c,T)=\text{P}_R\int_{0}^{t_c}|\alpha(t)|^2dt+\int_{t_c}^{T}\frac{1}{2}(\text{P}_{R}+\text{P}_{d})|\alpha(t)|^2 dt.
\end{multline}

We consider a square pulse of length $T$ and constant intensity $|\alpha|^2$ such that  $|\alpha|^2 T=\bar{n}$. Combining all these results and using Eq. (\ref{fidelity}) and $\rho_{10}^{(c)}(t_c)=\rho_{01}^{(c)\ast}(t_c)$ gives us a $(t_{c},T)$-dependent fidelity:
\begin{equation}
\label{fidtime}
F(t_c,T) =\frac{1}{2}+\frac{1}{2}e^{-\gamma(t_c,T)}\frac{\sqrt{\mathcal{N}}}{\mathcal{D}(t_{c})}\cos{\phi(t_c,T)}
\end{equation}
where we have defined 
\begin{eqnarray}
\frac{\mathcal{N}}{\Gamma^{4}} & = &\left (\frac{4\delta^2}{\Gamma^{2}}+(1-\beta_{0})^2\right)\beta_{0}\beta_{1},\\
\mathcal{D}(t_{c}) & = &(1/2)(4\delta^2+\Gamma^2)(1-\text{P}_{sc}e^{-\text{P}_R |\alpha|^2 t_c})\\
\phi(t_c,T) & =&\phi_z+\omega_{01}'(T-t_c)+\text{arctan}\left[\frac{2\delta/\Gamma}{(1-\beta_{0})}\right]
\end{eqnarray}
with $\phi_z=\omega_{01}(\text{z}-\text{z}_0)/v_\text{R}$,  $\omega_{01}'=\omega_{01}+4\beta_{0} |\alpha|^2\delta/\Gamma/(\delta^2/(\Gamma/2)^2+1)$ and $\gamma(t_c,T)=|\alpha|^2(\text{P}_R (t_c+T)/2+\text{P}_{d}(T-t_c)/2)$

To elucidate the physics contained in the expression for the fidelity let us consider a specific case where $\Gamma_{0,1\text{D}} = \Gamma_{1,1\text{D}}$, $\Gamma'=0$, $\delta = 0$, $\phi_z=\omega_{01}(\text{z}-\text{z}_\text{R})/v_\text{R} = q\times2\pi$ with $q$ being an integer. On using these conditions in Eq. (\ref{fidtime}) we get
\begin{equation}
\label{fidt}
F(t_c,T)=\frac{1}{2}+\frac{1}{2}\left(\frac{e^{-\bar{n}/2 }}{2-e^{-\frac{\bar{n}}{2} t_c/T}}\right)\cos\left(\omega_{01}T\left[1-\frac{t_c}{T}\right]\right).
\end{equation}
Note that in deriving the expression for fidelity, we have assumed the detector efficiency $\eta$ to be small so that the probability of detecting a photon is small. We plot the fidelity derived in Eq. (\ref{fidt}) for $T|\alpha|^2 = \bar{n} = 0.8$ and $\omega_{01} = 5\frac{2\pi}{T}$, as a function of $t_c/T$, in Fig. \ref{fidplot} (a). We find that the fidelity oscillates depending on the time of the click (detection of a photon) and that, for the given conditions, the amplitude decays with time. This is because, at later detection times, there is a larger probability that the emitter has already decayed, and hence the transmission is dominated by the direct transmission (the unity term in Eq. (\ref{eq73}) ). This does not create a superposition and hence the fidelity becomes lower.

In Fig. \ref{fidplot} (b) we plot the fidelity as a function of $\beta$, $(\beta = \beta_{0}+\beta_{1})$ assuming $\beta_{0} = \beta_{1}$ for different coherent-pulse average photon numbers. Note that $F_+ = 1-F_-$ where $F_{\pm}=|\langle \Psi^{\pm}| \Psi \rangle|^2$, so the fidelity for the symmetric superposition state $|\Psi^+\rangle = (|0\rangle + |1\rangle)/\sqrt{2}$ is equal to the fidelity with respect to the antisymmetric state  $|\Psi^-\rangle$  mirrored about $F=1/2$. In an experiment, the time of detection $t_c$ is randomly distributed according to the intensity (\ref{intensityRef}), and as such doing many of these experiments would on average yield a fidelity $\bar{F}=\int_{0}^{T}I_{out}(t_{c})F(t_c,T) dt_{c}/\int_{0}^{T}I_{out}(t_c)dt_{c}$, if we do not condition on a particular detection time. Taking the average results in 
\begin{align}
\bar{F}&=\frac{1}{2}+\frac{1}{2}\frac{\sin(\omega_{01 }T)}{\omega_{01 }T} \frac{e^{-\bar{n}/2}}{2-e^{-\bar{n}/2}}.
\end{align}
For suitable limits this can be simplified to 
\begin{align}
\bar{F}&\approx\frac{1}{2-e^{-\bar{n}/2}} \hspace{47pt}\text{for}\hspace{10pt} \omega_{01}\ll\frac{2\pi}{T}\\
\bar{F}&\approx\frac{1}{2}+\frac{1}{2}\frac{\sin(\omega_{01 }T)}{\omega_{01 }T} \hspace{12pt}\text{for}\hspace{26pt} \bar{n}\ll1.
\end{align}
From this we find, e.g., for $\omega_{01}\ll\frac{2\pi}{T}$, $\bar{F}\approx0.7$ for $\bar{n}=1$ and $\bar{F}\approx(1-\bar{n}/2)$ for $\bar{n}\ll1$. In the limit of $\omega_{01}\gg\frac{2\pi}{T}$ the fidelity reaches a value for a completely mixed state of $F=1/2$. This result is an instance of Heisenberg's `energy-time' uncertainty of the $\Lambda$-system state. If the detection-time interval is sufficiently short we cannot resolve the frequency resulting in a superposition of the possible outcomes. Furthermore, the fidelity decreases with a larger number of photons in the input coherent pulse because the state will have a larger decoherence due to scattering of additional photons.  
\section{Summary}
We have developed a theoretical framework for solving photon scattering from multiple scatterers in a $1$D waveguide. The formalism can be applied to any system of multi-level quantum emitters coupled to a $1$D waveguide mode. We have explicitly demonstrated how to apply the formalism to single-photon/weak-coherent pulse scattering. Our formalism conveniently employs the method of the effective operators to solve the possibly complicated dynamics of the emitters arising from the interaction with the incoming photons. Our approach is applicable to both single and double-sided waveguides and can also include chirality in the coupling. We have shown with several generic examples how one can apply the developed photon-scattering relation to experimentally viable physical systems. In particular, we show how our photon scattering formalism gives a direct solution to the nontrivial problem of generation of a superposition state based on detection of scattered photons.

It is worth emphasizing that this is a general framework that can be applied in many different contexts. The examples are therefore mainly meant as an illustration of how to apply the technique to achieve non-trivial results with limited calculations. In particular, we have already applied the formalism to describe entanglement generation between distant emitters in Ref. \cite{DasPRL17}. Such protocols may play an important role in future emerging quantum technologies. In this context, waveguides are particular useful for distributing information and we see wide application of our formalism both for optical and microwave qubits.
\begin{acknowledgments}
SD, VE, and AS gratefully acknowledge financial support from ERC Grant QIOS (Grant No. 306576) and the Danish Council for Independent Research (Natural Science). 
FR gratefully acknowledges financial support from the Humboldt Foundation. 
\end{acknowledgments}
\appendix
\begin{widetext}
\section{Detailed derivation of the photon-scattering relation}
In this appendix we provide a detailed derivation of the photon-scattering relation Eq. (\ref{eq5}) between the amplitudes of the incoming and outgoing photons. We start by substituting Eq. (\ref{eq3}) into Eq. (\ref{eq2}) and then comparing the RHS and LHS of Eq. (\ref{eq2}) to get
\bea
\label{eqa1}
&&i\sum_{k_{f}}\sqrt{\frac{\hbar\omega_{k_{f}}}{2}}\vec{F}_{k_{f}}(\vec{r}_{\perp})\hat{a}_{k_{f}}e^{i(k_{f}\text{z}-\omega_{k_{f}}t)} = i\int d\vec{r'_{\perp}}\mathbf{G}_{f}(\vec{r}_{\perp},t,\vec{r'}_{\perp},0)\epsilon(\vec{r'}_{\perp})\sum_{k_{f}}\sqrt{\frac{\hbar\omega_{k_{f}}}{2}}\vec{F}_{k_{f}}(\vec{r'}_{\perp})\hat{a}_{k_{f}}e^{ik_{f}\text{z}}\nonumber\\
&+&\left(\frac{i\omega}{2\hbar}\right)\sum_{jj'}\sum_{gg'}\int^{\infty}_{0} d\tau' e^{i\omega_{gg'}\tau}\hat{\sigma}_{g'g}\mathbf{G}_{f}(\vec{r}_{\perp},t,\vec{r}_{j\perp},t')\sum_{ee'}\left[\vec{d}^{j}_{ge}(\tilde{\mathcal{H}}_{\text{nh}})^{-1}_{ee'}\vec{d}^{j'}_{e'g}\right]\nonumber\\
&\times&\int d\vec{r'}_{\perp}\bigg[\mathbf{G}_{f}(\vec{r}_{j'\perp},t',\vec{r'}_{\perp},0)\epsilon(\vec{r'}_{\perp})i\sum_{k_{f}}\sqrt{\frac{\hbar\omega_{k_{f}}}{2}}\vec{F}_{k_{f}}(\vec{r'}_{\perp})\hat{a}_{k_{f}}e^{ik_{f}\text{z}}+\mathbf{G}_{b}(\vec{r}_{j'\perp},t',\vec{r'}_{\perp},0)\epsilon(\vec{r'}_{\perp})\nonumber\\
&&i\sum_{k_{b}}\sqrt{\frac{\hbar\omega_{k_{b}}}{2}}\vec{F}_{k_{b}}(\vec{r'}_{\perp})\hat{a}_{k_{b}}e^{ik_{b}\text{z}}\bigg]+\mathcal{F}
\eea
\bea
\label{eqa2}
&&i\sum_{k_{b}}\sqrt{\frac{\hbar\omega_{k_{b}}}{2}}\vec{F}_{k_{b}}(\vec{r}_{\perp})\hat{a}_{k_{b}}e^{i(k_{b}\text{z}-\omega_{k_{b}}t)} = i\int d\vec{r'_{\perp}}\mathbf{G}_{b}(\vec{r}_{\perp},t,\vec{r'}_{\perp},0)\epsilon(\vec{r'}_{\perp})\sum_{k_{b}}\sqrt{\frac{\hbar\omega_{k_{b}}}{2}}\vec{F}_{k_{b}}(\vec{r'}_{\perp})\hat{a}_{k_{b}}e^{ik_{b}\text{z}}\nonumber\\
&+&\left(\frac{i\omega}{2\hbar}\right)\sum_{jj'}\sum_{gg'}\int^{\infty}_{0} d\tau' e^{i\omega_{gg'}\tau}\hat{\sigma}_{g'g}\mathbf{G}_{b}(\vec{r}_{\perp},t,\vec{r}_{j\perp},t')\sum_{ee'}\left[\vec{d}^{j}_{ge}(\tilde{\mathcal{H}}_{\text{nh}})^{-1}_{ee'}\vec{d}^{j'}_{e'g}\right]\nonumber\\
&\times&\int d\vec{r'}_{\perp}\bigg[\mathbf{G}_{f}(\vec{r}_{j'\perp},t',\vec{r'}_{\perp},0)\epsilon(\vec{r'}_{\perp})i\sum_{k_{f}}\sqrt{\frac{\hbar\omega_{k_{f}}}{2}}\vec{F}_{k_{f}}(\vec{r'}_{\perp})\hat{a}_{k_{f}}e^{ik_{f}\text{z}}+\mathbf{G}_{b}(\vec{r}_{j'\perp},t',\vec{r'}_{\perp},0)\epsilon(\vec{r'}_{\perp})\nonumber\\
&&i\sum_{k_{b}}\sqrt{\frac{\hbar\omega_{k_{b}}}{2}}\vec{F}_{k_{b}}(\vec{r'}_{\perp})\hat{a}_{k_{b}}e^{ik_{b}\text{z}}\bigg]+\mathcal{F}
\eea
The symbol $\mathcal{F}$ here stands for noise which corresponds to the field not into the waveguide mode and can be expressed in terms of $\mathcal{E}_{\text{rest},\zeta}(\vec{r},t)$ and the Green's function $\mathbf{G}_{\text{rest},{\zeta}}(\vec{r},t, \vec{r'},t')$. We next solve the space and time integrals in Eq. (\ref{eqa1}) and (\ref{eqa2}) and convert the sum to an integral $\sum_{k} \rightarrow \frac{1}{\sqrt{2\pi}}\int dk$.  Finally after multiplying both sides with the mode function $\epsilon\vec{F}^\ast_{k_{\zeta}}(\vec{r}_\perp)$, integrating over the transverse plane and on comparing the terms on the RHS and LHS, we arrive at an input-output formalism between the incoming and scattered photons represented respectively by the mode operators, $a_{o,f}$ and $a_{in,f}$   
\bea
\label{eqa3}
a_{o,f}\left(t-\frac{\text{z}}{v_{g}}\right) & = & a_{in,f}\left(t-\frac{\text{z}}{v_{g}}\right) +\left(\frac{i\omega_{0}\pi}{\hbar v_{g}}\right)\sum_{jj'}\sum_{gg'}e^{-i\omega_{gg'}|\text{z}-\text{z}_{j}|/v_{g}}\hat{\sigma}_{g'g}\sum_{ee'}\bigg[\left(\mathcal{A}^{\ast jf}_{ge}(H_{\text{nh}})^{-1}_{jj'}\mathcal{A}^{j'f}_{e'g}\right)\nonumber\\
& &a_{in,f}(0)+\left(\mathcal{A}^{\ast jf}_{ge}(H_{\text{nh}})^{-1}_{ee'}\mathcal{A}^{j'b}_{e'g}\right)e^{-2i\vec{k}_{0}\text{z}_{j}}a_{in,b}(0)\bigg]+\mathcal{F}
\eea
\bea
\label{eqa4}
a_{o,b}\left(t+\frac{\text{z}}{v_{g}}\right) & = & a_{in,b}\left(t+\frac{\text{z}}{v_{g}}\right) +\left(\frac{i\omega_{0}\pi}{\hbar v_{g}}\right)\sum_{jj'}\sum_{gg'}e^{-i\omega_{gg'}|\text{z}-\text{z}_{j}|/v_{g}}\hat{\sigma}_{g'g}\sum_{ee'}\bigg[\left(\mathcal{A}^{\ast jf}_{ge}(H_{\text{nh}})^{-1}_{ee'}\mathcal{A}^{j'f}_{e'g}\right)\nonumber\\
& &a_{in,b}(0)+\left(\mathcal{A}^{\ast jf}_{ge}(H_{\text{nh}})^{-1}_{jj'}\mathcal{A}^{j'b}_{e'g}\right)e^{2i\vec{k}_{0}\text{z}_{j}}a_{in,f}(0)\bigg]+\mathcal{F}
\eea
Here $f (b)$ signifies the forward (backward) direction of propagation for the incoming and scattered photons. Note that we consider both the forward and backward contributions to the input field as well as the scattered fields as we assume a double-sided waveguide with input possible from both ends. In deriving the above set of equations, we have expanded $\omega_{k,f/b} = \omega_{0}+v_{g_{f/b}}(k_{f/b}-k_{0})$ with $k_{f/b} = \pm k$. Furthermore, we have written the Green's function in terms of the mode function and assumed that the transverse field into the waveguide have the mode functions of the form $\vec{F}_{k_{f}}(\vec{r}_{\perp}) = \vec{F}_{k_{f}}(\vec{r}_\perp)e^{i\vec{k}_{f}\text{z}}$, $\vec{F}_{k_{b}}(\vec{r}_\perp) = \vec{F}_{k_{b}}(\vec{r}_{\perp})e^{i\vec{k}_{b}\text{z}}$. The coupling strength $\mathcal{A}^{j, (f/b)}_{eg}$ in the above photon-scattering relation is defined as a product of the emitter's dipole moments and the field-mode function in the form $\mathcal{A}^{j(f/b)}_{eg} =  \sqrt{\frac{\pi\omega}{\hbar v_{g}}}\left[\vec{d}^{j}_{eg}\cdot\vec{F}_{k_\zeta}(r_{j_\perp})\right]$. Finally we have also defined different forward and backward mode operators of the incoming and scattered field as
\bea
\label{eqb5}
\hat{a}_{o,f/b}\left(t-\frac{\text{z}}{v_{g}}\right) & = &\sqrt{\frac{v_{g}}{2\pi}} \int~dk_{f/b}e^{-i\delta k_{f/b}v_{g}(t-\frac{\text{z}}{v_{g}})}\hat{a}_{k_{f/b}}\\
\hat{a}_{in,f/b} (t)& = &\sqrt{\frac{v_{g}}{2\pi}} \int~dk_{f/b}e^{-i\delta k_{f/b}v_{g}t}\hat{a}_{k_{f/b}}\\
\eea
Eq. (\ref{eq5}) and Eq. (\ref{eq6}) then follows from Eq. (\ref{eqa3}) and Eq. (\ref{eqa4}) with the decay into the forward and backward modes of the waveguide $\Gamma^{(f/b)}_{eg}$, defined in terms of the coupling strengths $\mathcal{A}^{(f/b)}_{eg}$ and their complex conjugate.
\end{widetext}
\begin{widetext}
\section{Derivation of the waveguide-mediated coupling between emitters}
The waveguide-mediated decay and shifts of the emitter's excited state are given by, 
\bea
\label{eqb1}
\Gamma^{jj',e'e}_{gg'} & = &\frac{2\omega_{e'g'}^{2}}{\hbar v^2_{g}}\left\{\vec{d}^{j}_{e'g}\cdot\mathbf{Im}\overleftrightarrow{\mathbf{G}}_\zeta(\vec{r}_{j},\vec{r}_{j'},\omega_{e'g'})\cdot\vec{d}^{j'}_{g'e}\right\},\\
\label{eqb2}
\Omega^{jj',e'e}_{gg'} & = &\mathbf{P}\int d\omega \left(\frac{\omega^2}{\hbar\pi v^2_{g}}\right)\bigg\{\frac{\vec{d}^{j}_{e'g}\cdot\mathbf{Im}\overleftrightarrow{\mathbf{G}}_\zeta\cdot\vec{d}^{j'}_{g'e}}{(\omega-\omega_{e'g'}+i\epsilon)}\bigg\}.
\eea
\begin{figure}
	\includegraphics[height = 5 cm]{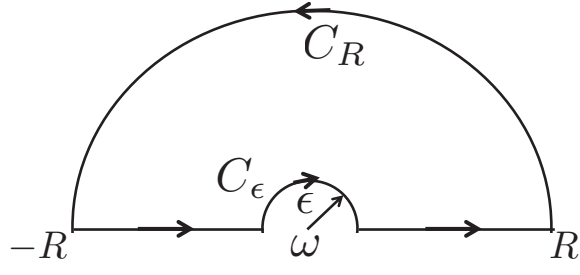}
	\caption{Contour for evaluating the principal-value integral}
	\label{config}
\end{figure}
Now considering the expression for $\mathbf{Im}\overleftrightarrow{\mathbf{G}}_\zeta(\vec{r}_{j},\vec{r}_{j'},\omega_{e'g'})$ in Eq. (\ref{eq10}) and substituting it into the above Eqs. (\ref{eqb1}) and (\ref{eqb2}) we get,
\bea
\label{eqb3}
\Gamma^{jj',e'e}_{gg'} & = &2\sum_{\zeta}\mathcal{A}^{j\zeta}_{k}\mathcal{A}^{\ast j'\zeta}_{k}\cos\left(k_\zeta|\text{z}_{j}-\text{z}_{j'}|\right),\\
\label{eqb4}
\Omega^{jj',e'e}_{gg'} & = &\frac{1}{2\hbar v_{g}}\sum_{\zeta}\mathbf{P}\int^{\infty}_{-\infty} d\omega' ~\omega' (g^{j\zeta}_{\omega'/v_{g}}g^{\ast j'\zeta}_{\omega'/v_{g}})\left[\frac{\cos\left(\omega'|\text{z}_{j}-\text{z}_{j'}|/v_{g}\right)}{(\omega'-\omega+i\epsilon)}\right]
\eea
where $g^{j\zeta}_{\omega'/v_{g}} = \vec{d}^{j}_{eg}\cdot\vec{F}_{\omega'/v_{g}}(\vec{r}_{j\perp})$. We next expand the cosine term in the above integral as $[\exp(i\omega'|\text{z}_{j}-\text{z}_{j'}|/v_{g})+\exp(-i\omega'|\text{z}_{j}-\text{z}_{j'}|/v_{g})]/2$ and write Eq. (\ref{eqb4}) as sum of two integrals. We then solve the integral with the positive frequency integrand by the method of Cauchy's principal value over the contour shown in Fig. (\ref{config}). It can be seen clearly that the integral does not have a pole inside the big contour $C_{R}$. Hence from the residue theorem, we find that the total integral $\left[\int_{C_{R}} + \int^{\omega-\epsilon}_{-R} +\int_{C_{\epsilon}}+ \int_{\omega+\epsilon}^{R}\right]d\omega~f(\omega) = 0$. However, this can be rewritten as $\left[\int_{C_{R}} + \int^{\omega-\epsilon}_{-R} + \int_{\omega+\epsilon}^{R}\right]d\omega~f(\omega) = -\int_{C_{\epsilon}}d\omega~f(\omega)$. Thus, in the limit of $R\rightarrow\infty$ the right hand side can be evaluated in terms of the value of the analytical function $f(\omega)$ for the small contour $C_\epsilon$.  On evaluating the small contour $C_{\epsilon}$ we get $\int^{\infty}_{-\infty} d\omega f(\omega') = -i\pi f(\omega)$, where $f(\omega') = \omega' (g^{j\zeta}_{\omega'/v_{g}}g^{\ast j'\zeta}_{\omega'/v_{g}})e^{i\omega'|z_{j}-z_{j'}|/v_{g}}$. The integral for the negative frequency integrand $\exp(-i\omega'|\text{z}_{j}-\text{z}_{j'}|/v_{g})/2$ can be solved similarly by choosing a contour that is mirror reflection of Fig. (\ref{config}) about the real axis. This then gives for the small contour $C_\epsilon$, that goes counter-clockwise $\int^{\infty}_{-\infty} d\omega f(\omega') = i\pi f(\omega)$, where now $f(\omega') = \omega' (g^{j\zeta}_{\omega'/v_{g}}g^{\ast j'\zeta}_{\omega'/v_{g}})e^{-i\omega'|z_{j}-z_{j'}|/v_{g}}$. Finally, on substituting the evaluated integral into Eq. (\ref{eqb4}) we find the principal-value integral to be 
\bea
\label{eqb5}
\Omega^{jj',e'e}_{gg'} = -\sum_{\zeta}\mathcal{A}^{j\zeta}_{k}\mathcal{A}^{\ast j'\zeta}_{k}\sin\left(k_\zeta|z_{j}-z_{j'}|\right),
\eea
where we have used the definition of $\mathcal{A}^{j\zeta}_{k}$ from Sec. III. The evaluated integral thus gives  Eq. (\ref{eq11}) and Eq. (\ref{eq12}) of Sec. III. 
\end{widetext}
\begin{widetext}
\section{Definition of the effective detuning and rates for the two-emitter system}
In this appendix we define the effective detunings and decay rates introduced as a part of the non-Hermitian Hamiltonian in Eq. (\ref{eq62}) for the two-emitter system with one being a two-level system while the other system is a three-level in V-configuration. 
\begin{subequations}
\begin{align}
	\delta_{1,\text{eff}}^{-1}&\equiv\left[\tilde{\delta}_1+\frac{\Gamma_{12}^2}{4\tilde{\delta}_2}+\frac{\Gamma_{13}^2}{4\tilde{\delta}_3}-\frac{(\Omega-i\Gamma_{23})(\frac{\Gamma_{12}^2}{4\tilde{\delta}_2}+\frac{\Gamma_{13}^2}{4\tilde{\delta}_3})-\Gamma_{12}\Gamma_{13}}{(\Omega-i\Gamma_{23})-\frac{4\tilde{\delta}_2 \tilde{\delta}_3}{\Omega-i\Gamma_{23}}}\right]^{-1}\\
	\delta_{2,\text{eff}}^{-1}&\equiv\left[\tilde{\delta}_2+\frac{\Gamma_{12}^2}{4\tilde{\delta}_1}-\frac{(\Omega-i\Gamma_{23})^2}{4\tilde{\delta}_3}-\frac{\Gamma_{13}(\frac{\Gamma_{12}^2}{4\tilde{\delta}_1}-\frac{(\Omega-i\Gamma_{23})^2}{4\tilde{\delta}_3})+\Gamma_{12}(\Omega-i\Gamma_{23})}{\Gamma_{13}-\frac{\tilde{4\delta}_1 \tilde{\delta}_3}{\Gamma_{13}}}\right]^{-1}\\
	\delta_{3,\text{eff}}^{-1}&\equiv\left[\tilde{\delta}_3+\frac{\Gamma_{13}^2}{4\tilde{\delta}_1}-\frac{(\Omega-i\Gamma_{23})^2}{4\tilde{\delta}_2}-\frac{\Gamma_{12}(\frac{\Gamma_{13}^2}{4\tilde{\delta}_1}-\frac{(\Omega-i\Gamma_{23})^2}{4\tilde{\delta}_2})+\Gamma_{13}(\Omega-i\Gamma_{23})}{\Gamma_{12}-\frac{4\tilde{\delta}_1 \tilde{\delta}_2}{\Gamma_{12}}}\right]^{-1}
	\end{align}
\end{subequations}		
\begin{subequations}	
\begin{align}
	\Gamma_{12,\text{eff}}^{-1}&\equiv\left[-\frac{i}{2}\Bigg(\Gamma_{12}+\frac{4\tilde{\delta}_1\tilde{\delta}_2}{\Gamma_{12}}+\frac{\frac{\Gamma_{13}^2}{\Gamma_{12}}\tilde{\delta}_2-4\frac{(\Omega/2-i\Gamma_{23}/2)^2}{\Gamma_{12}}\tilde{\delta}_1-(\Omega/2-i\Gamma_{23}/2)\Gamma_{13}(1-4\frac{\tilde{\delta}_1\tilde{\delta}_2}{\Gamma_{12}^2})}{\tilde{\delta}_3-\frac{(\Omega/2-i\Gamma_{23}/2)\Gamma_{13}}{\Gamma_{12}}}\Bigg)\right]^{-1}\\
	 \Gamma_{13,\text{eff}}^{-1}&\equiv\left[-\frac{i}{2}\Bigg(\Gamma_{13}+\frac{4\tilde{\delta}_1\tilde{\delta}_3}{\Gamma_{13}}+\frac{\frac{\Gamma_{12}^2}{\Gamma_{13}}\tilde{\delta}_3-4\frac{(\Omega/2-i\Gamma_{23}/2)^2}{\Gamma_{13}}\tilde{\delta}_1-(\Omega/2-i\Gamma_{23}/2)\Gamma_{12}(1+\frac{\tilde{\delta}_1\tilde{\delta}_3}{\Gamma_{13}^2})}{\tilde{\delta}_2-\frac{(\Omega/2-i\Gamma_{23}/2)\Gamma_{12}}{\Gamma_{13}}}\Bigg)\right]^{-1}\\
	\Gamma_{23,\text{eff}}^{-1}&\equiv\left[(\Omega/2-i\Gamma_{23}/2)-\frac{\tilde{\delta}_{2}\tilde{\delta}_{3}}{\Omega/2-i\Gamma_{23}/2}+\frac{1}{4}\frac{\left(\Gamma_{12}-\frac{\Gamma_{13}\tilde{\delta}_{2}}{\Omega/2-i\Gamma_{23}/2}\right)\left(\Gamma_{13}-\frac{\Gamma_{12}\tilde{\delta}_{3}}{\Omega/2-i\Gamma_{23}/2}\right)}{\tilde{\delta}_{1}+\frac{1}{4}\frac{\Gamma_{12}\Gamma_{13}}{\Omega/2-i\Gamma_{23}/2}}\right]^{-1}.
	\end{align}
\end{subequations}
\end{widetext}


\end{document}